\documentclass[prb,aps,twocolumn]{revtex4}
\usepackage{graphicx}
\usepackage{amsmath}
\usepackage{color}
\def\blue{\color{blue}}

\newcommand{\beq}{\begin{equation}}
\newcommand{\eeq}{\end{equation}}
\newcommand{\bea}{\begin{eqnarray}}
\newcommand{\eea}{\end{eqnarray}}
\newcommand{\e}{\varepsilon}
\newcommand{\up}{\uparrow}
\newcommand{\down}{\downarrow}

\newcommand{\bk}{{\vec k}}
\newcommand{\bp}{{\vec p}}
\newcommand{\bq}{{\vec q}}

\newcommand{\bse}{\begin{subequations}}
\newcommand{\ese}{\end{subequations}}
\newcommand{\bwt}{\begin{widetext}}
\newcommand{\ewt}{\end{widetext}}

\newcommand{\bsu}{\begin{subequations}}
\newcommand{\esu}{\end{subequations}}

\graphicspath{{./figures/}}

\begin{document}

\title{Collective modes in superconductors with competing $s$- and $d$-wave interactions}
\author{Saurabh Maiti$^{1,2}$ and   P. J. Hirschfeld$^{1}$ }
\affiliation {~$^1$Department of Physics, University of Florida, Gainesville, FL 32611}
\affiliation {~$^2$National High Magnetic Field Laboratory, Tallahassee, FL 32310}
\date{\today}

\begin{abstract}
We calculate the collective mode spectrum in models of superconductors with attractive interactions in $s$-and $d$ channels as a function of their relative strength, across a phase diagram that includes transition between $s$, $s+id$, and $d$-wave ground states.  For one-band systems, we recover the largely known results for the phase, amplitude, and Bardasis-Schrieffer modes of pure $s$ or $d$-states, and show how the well-defined Bardasis-Schrieffer mode softens near the $s+id$ phase boundary and evolves in a characteristic manner  through the $s+id$ phase as a mixed symmetry mode.
For two-band systems, we consider a model of hole-doped Fe-based superconductors,
and find in the case of an $s$-wave ground state a
well-defined Bardasis-Schrieffer mode below the lowest gap edge,
as well as a second, damped mode of this type between the two gap energies.  Both
modes soften as the $s+id$ phase is approached,
and only a single ``mixed-symmetry Bardasis-Schrieffer mode" below the pairbreaking
continuum propagates in the $s+id$ phase itself.  These modes coexist with a damped
Leggett mode with collective frequency between the two
gap scales.  In the pure $d$-state, no Bardasis-Schrieffer type $s$ excitonic mode exists
at low T.  We briefly discuss Raman scattering experiments and how they can be
used to identify an $s+id$ state, and to track the evolution of competing
$s$- and $d$- interactions in these systems.
\end{abstract}


\maketitle
\section{Introduction}
In contrast to cuprate superconductors, which are believed to have universal $d$-wave symmetry,\cite{ScalapinoReview95,Tsuei00} the Fe-based superconductors (FeSC) appear to manifest generic $s$-wave pairing, with an order parameter that likely changes sign between Fermi surface (FS) sheets.\cite{HKMReview11,ChubukovReview12,KurokiHosonoReview15}  On the other hand, calculations of pairing by exchange of spin fluctuations have suggested from the early days of research on these materials that the $d$-wave channel can be strongly competitive, and might under some circumstances become the dominant pair symmetry.\cite{Kuroki08,Graser09}  Within an RPA treatment, it was argued that overdoping either by holes or electrons away from a 6 electrons/Fe parent material should lead to a $d$-wave ground state.\cite{Maiti11a,Maiti11b} This gave rise to the possibility that the symmetry broken $s$-wave phase could make a low-temperature transition to a $d$-wave phase, and several authors argued that this should proceed through an intervening $s+id$ state, in which $s$ and $d$ wave functions are combined with fixed relative phase $\pi/2$ and time reversal symmetry $\cal T$ is broken.\cite{Hanke_sid,Chubukov_sid}   While such a transition would be of great potential interest, and represent the first example of its kind, an $s+id$ state is not trivial to detect.   Although $\cal T$ is broken, the state is not chiral, and thus does not manifest spontaneous edge currents as discussed, e.g. in the context of the $p+ip'$ state of Sr$_2$RuO$_4$.   Its quasiparticle excitations are fully gapped, but so are those of the $s$ wave state out of which it evolves, so thermodynamic signatures of the transition are likely to be weak.

The proximity of different pairing channels is an unusual situation in superconductors(SC), but was studied quite early in the  pioneering work of Bardasis and Schrieffer,\cite{BardasisSchrieffer61} where the effect of fluctuations in a subdominant pairing channel was investigated in a conventional $s$-symmetry ground state. The motion of the order parameter was found to include a collective mode corresponding to the oscillation of the phase of the subdominant pairing channel,  with the $q\rightarrow 0$ frequency dependent on the difference between the inverse of the two pairing interaction components. This frequency is
located below the pair-breaking edge of the condensed $s$-wave system.  This ``Bardasis-Schrieffer (BS)" mode (sometimes referred to as a particle-particle exciton) was never convincingly observed in conventional superconductors owing, presumably, to a dearth of systems exhibiting a strong $s$-$d$ competition. Taking a hint from the proximity of $s$ and $d$ channels predicted by spin fluctuation theory, FeSC can be  viewed as excellent candidates to probe such modes. A search using Raman scattering was proposed in FeSC by Devereaux and Scalapino.\cite{DevereauxScalapino09} Recently, two electronic Raman measurements on Ba$_{1-x}$K$_x$Fe$_2$As$_2$\cite{Hackl12,Hackl14} and NaFe$_{1-x}$Co$_x$As,\cite{Blumberg14} found features associated with BS modes. The exact identification of these features with a BS mode is hindered by the fact that these systems posses multiple gaps and the exact nature of possible collective modes and their evolution across a typical doping phase diagram is not clearly known. Nevertheless, these discoveries raise the prospect of systematic studies of the interaction strengths and collective modes in different channels in FeSC for the first time.

To facilitate the interpretation of experimental results in this area as they develop, we provide here the collective mode spectrum in a simplified model of an unconventional superconductor with competing pairing in $s$ and $d$ channels. The prime focus of this work is the study of collective modes in a non-$s$ wave ground state. We thus explicitly account for the possibility of a transition to an $s+id$ state between pure $s$ and $d$ phases and investigate collective modes in the $s+id$ state and report a mixed symmetry collective mode which couples the amplitude and phase sectors of the fluctuations in the SC order parameter and exhibits oscillations in both $s$ and $d$ symmetry channels. As a demonstration of our approach, we also study the simpler, one-band case, and drive it through $s$ to $s+id$ to $d$ transitions. We reproduce the well known results for collective modes in single band superconductor and use them as a benchmark to discuss the differences in a multiband system. We demonstrate the existence of the mixed symmetry collective mode in both one and multi-band cases.

The study of collective modes in `exotic' non-pure-$s$ wave superconductors has some history:  Hirschfeld et al\cite{HirschfeldBphase} studied the analog of the $^3$He-B ``squashing modes" in the Balian-Werthamer $p$-wave ground state, discussing how such modes could be observed in optical conductivity.  Wu and Griffin\cite{WuGriffin} studied the possibility of an $s$-excition for systems with $d$-wave ground states, and showed that this mode does not propagate at low $T$. More recently, in the context of the FeSC, Devereaux and Scalapino investigated the role of a BS particle-particle exciton on the Raman spectrum of an $s_\pm$ superconductor.  \cite{DevereauxScalapino09} Khodas et. al. studied Raman signature of this mode and the role of density fluctuations in Fe-selenides.\cite{Khodas}  Bittner \emph{et. al.} derived general expressions for the collective modes in the $s$-wave ground state for  noncentrosymmetric systems.\cite{Einzel15}

Works on collective modes in $\mathcal{T}$-broken superconductors are somewhat rarer.  Several papers generalized the well-known collective modes of the $^3$He-A phase\cite{VW} to $p+ip'$ and $d+id'$ superconductors in situations where the two harmonics in question  corresponded to the basis functions of a 2D representation of the symmetry group.\cite{YipSauls,Hirschfeld92}  Balatsky \emph{et. al.}\cite{Balat} phenomenologically discussed a `clapping' type orbital  mode analogous to those discussed in Ref. \onlinecite{Hirschfeld92}, but for a general situation where the two harmonics $d$ and $d'$ were not necessarily degenerate. The same mode and its detection in Raman experiments was also discussed in Lee et. al.\cite{Zhang2}

This mode corresponded to an oscillation of the relative phase of the two components $d_{x^2-y^2}$ (d)  and $d_{xy}(d')$, two distinct representations of the tetragonal group. More recently, one of the authors\cite{SM_AVC} and Marciani \emph{et. al.}\cite{Benfatto} studied, in the context of FeSC, the Leggett modes (the oscillations of the relative phase between order parameters on two bands: see Ref \onlinecite{Leggett}) in a special $s$-wave $\mathcal{T}$ broken $s+is'$ SC where it was found that the mode softened at the boundaries of the $s+is'$ state (see also Refs. \onlinecite{Legget1,Legett2,Legett3}).

The current work is more along the spirit of the last work above in the sense that the collective modes in a multiband $s+id$ SC are investigated. The mixed symmetry collective mode that we find also softens at the boundaries of the $s+id$ state. Besides the mixed symmetry collective mode in the $s+id$ ground state, we also find, in the $s-$wave ground state, a damped Leggett mode (which has $s$-wave symmetry) residing between the multiple gaps in the system (in our language we do not consider them to be true collective modes of the system, but nevertheless appropriate response functions will show broad peaks). Additionally, besides the usual BS mode residing below the minimum gap in the system, we report another damped BS mode residing between the multiple gaps. Within our model, we can show that there is only one BS mode below the minimum gap for any interaction. The presence of multiple BS modes is therefore expected to be a generic feature of multiband systems.

In this work we have ignored the coupling to the density fluctuations. The main qualitative effect of ignoring the density fluctuations is related to the Bogoliubov-Anderson-Goldstone(BAG) mode. Although  the BAG mode is expected due to spontaneous breaking of U(1) symmetry during the SC transition, coupling to charge density fluctuations implies that the oscillation of the gauge degree of freedom is identical to the usual plasmon (``Anderson-Higgs mechanism").\cite{gauge} In conventional cases of 1-band and 2-band SC, it is well known that coupling to density fluctuations does not affect the mass of the Leggett\cite{Leggett,SM_AVC,Benfatto} or the BS modes.\cite{BardasisSchrieffer61} Since the arguments are based on gauge invariance and symmetries, we assume without proof that same hold for our multiband system. We expect our work to provide useful insight in terms of number of collective modes to be expected in a system and detecting a non-trivial multiband SC ground states ($s+id$ is our case). Although there are many ways to model the multiband scenario, we limit our considerations to a minimal model that can be readily applied to FeSC, as will be described later.

We study the collective modes by studying the possible excitations in the system within linear response in different angular momentum channels. This method is sometimes referred to as a generalized random phase approximation, and is known to yield results  identical to those  obtained from the kinetic equation method\cite{VW}.  We explicitly derive a 1-band case and extend the formula to the multiband scenario. We stress that the formulation has the great advantage of spitting out all the collective modes in a clean SC in all angular momentum channels with minimal effort. The biggest advantage is its scalability to multiband or multiorbital systems. The rest of the paper is organized as follows: In Sec. \ref{sec:1band} we specify our 1 band model, derive the collective mode equation and study the collective modes; reproducing the well known results. In Sec. \ref{sec:3poc} we discuss the modeling of a FeSC with 3 pockets, discuss the collective modes and highlight the differences with 1-band model. In Sec. \ref{sec:disc} we discuss our results in connection to FeSC and some of the recent Raman experiments. We summarize our main findings in Sec. \ref{sec:conclu}. The Appendix presents details of some of the calculations.

\section{Collective modes in a 1-band model}\label{sec:1band}
We revisit this simple model as this helps us in two ways: we can systematically tune the system through a $s$ to $s+id$ to $d$ transitions and trace the collective modes across the phase diagram (through the $s+id$ phase whose collective modes have not been addressed before); we will then use this result as the benchmark against which the multiband case will be compared. Since we are interested in studying the collective modes in the SC state, we only retain the interactions in the particle-particle (p-p) channel. As discussed in the introduction, the interactions in the particle-hole (p-h) channel (which couples the SC fluctuations to density fluctuations) will be dropped. The logic of our presentation will be the following: we start with an $s-$wave ground state; use the $d-$wave interaction as our tuning parameter to generate a phase diagram that scans through the $s$, $s+id$, $d$ wave regions; we then find the collective modes with both $s$ and $d$ symmetries in each region and consistently track them as the $d-$wave interaction is tuned.

\subsection{Model and phase diagram}
\begin{figure*}[htp]
$\begin{array}{ccc}
\includegraphics[width=0.6\columnwidth]{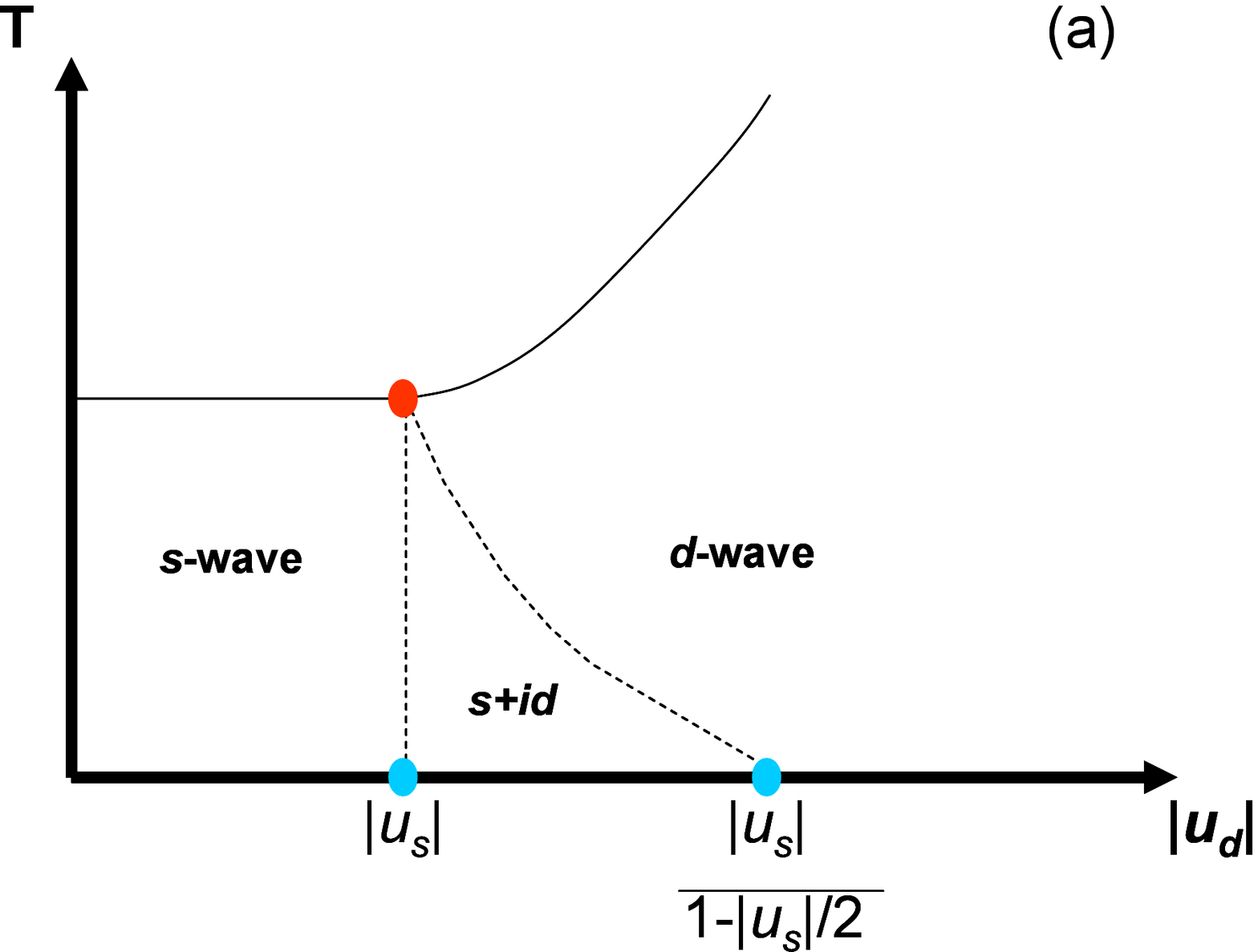}&
\includegraphics[width=0.7\columnwidth]{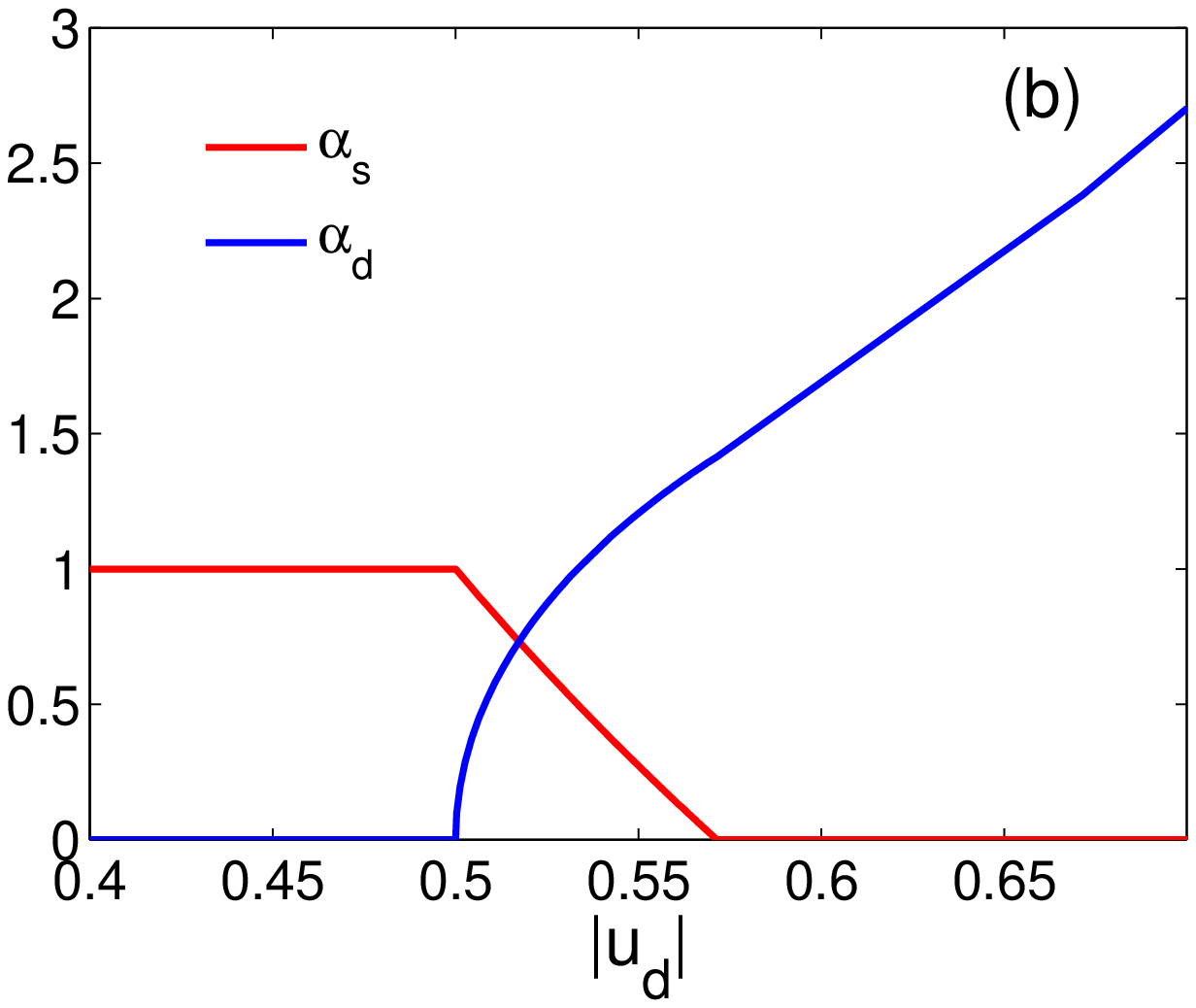}&
\includegraphics[width=0.7\columnwidth]{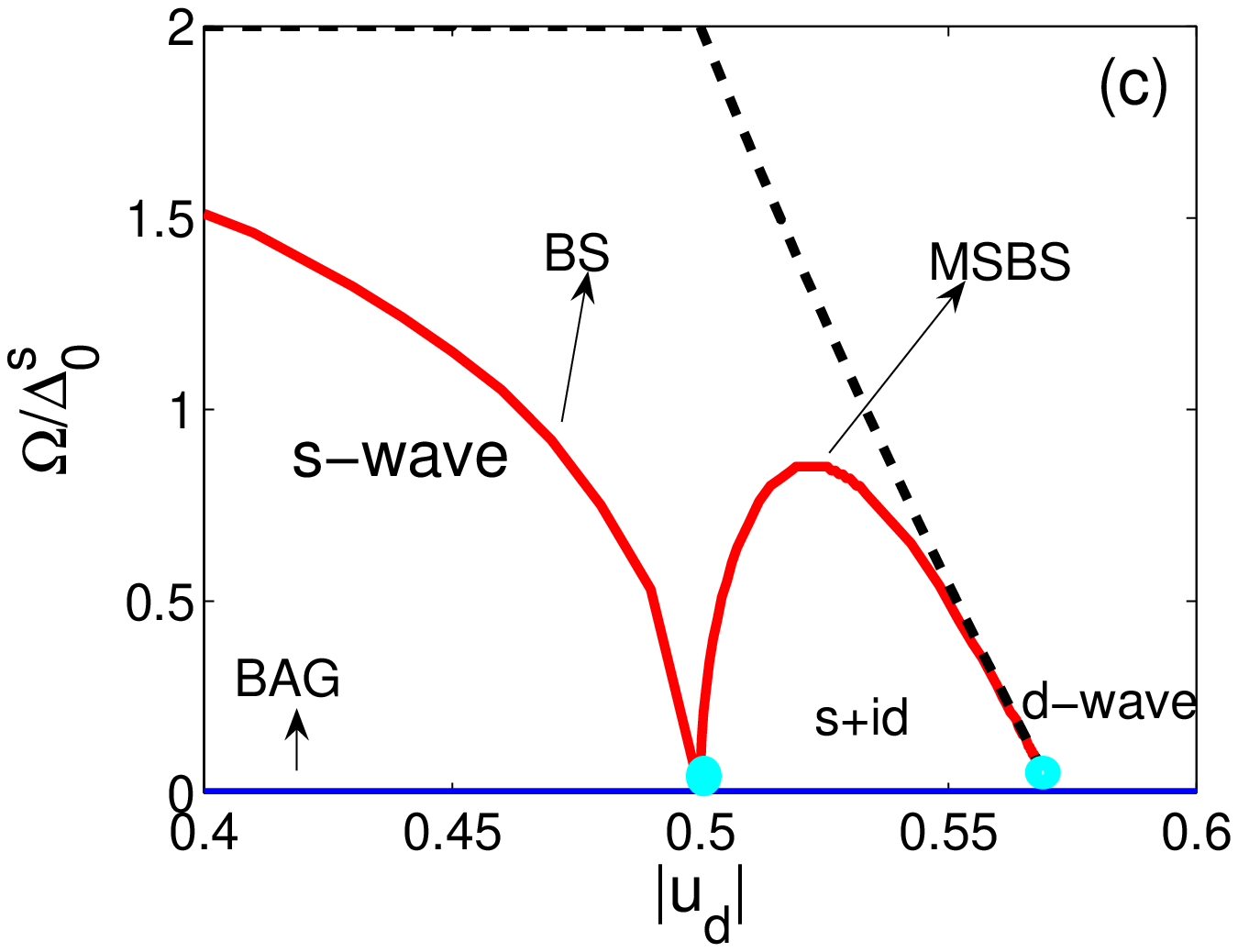}
\end{array}$\caption{
\label{fig:1} (a) The schematics of the phase diagram of the 1-band model. (b) The calculated evolution of the $s-$ and $d-$wave gaps $\alpha_s$ and $\alpha_d$ (normalized to the pure $s-$ wave gap) as a function of the $d-$wave interaction strength $u_d$. (c) The evolution of the collective modes(solid lines) in a 1-band model as $|u_d|$ is increased. The BS mode softens as the $s+id$ boundary(shown in light blue dots) and acquires a mixed symmetry (MS) character in the $s+id$ state and softens again at the other boundary. The dashed black line denotes the minimum gap in the system. Collective modes are well defined only below the minimum gap. There are no other damped resonances in this model. Here $u_s=-0.5$.}
\end{figure*}

Our one band model is a 2D Fermi liquid (FL) with the interaction $V(\bp,-\bp;\bk,-\bk)\equiv V(\bp,\bk)$ in the pairing channel. We decompose this interaction into different singlet angular momentum channels (limiting ourselves up to the $d-$wave harmonic):
\beq\label{eq:0}
V(\bp,\bk) = U^s + U^d f_{\bk}f_{\bp},
\eeq
where $f_{\bk}=\sqrt{2}\cos2\theta_{\bk}$. All the vectors are by definition on the circular fermi surface (FS) and $\theta_{\bk}$ is the angle of $\bk$ measured from the $k_x$-axis. We follow the convention where repulsion is denoted by the positive sign of $U$'s. We keep $U^s$ fixed and tune $U^d$. Within weak coupling, this results in the usual self consistency relation for the order parameter $\Delta_{\bp}$:\cite{SChriffer}
\beq\label{eq:1}\Delta_{\bp}=-\int_K~V(\bp,\bk)\frac{\Delta_{\bk}}{\omega_n^2 + \e_{\bk}^2 + |\Delta_{\bk}|^2},\eeq
where $\int_K$ stands for $T\sum_n \int \frac{d^2k}{(2\pi)^2}$. Taking as input that the only stable solutions are $s$, $s+id$, and $d$ states (no $s+d$),\cite{Chubukov_sid} and writing $\Delta_{\bp} = \Delta^s + \Delta^d f_{\bp}$ (with $\Delta^{s,d}$ as constants) we arrive at
\bea\label{eq:2}
\Delta^s&=&-U^s\int_K \frac{\Delta^s}{\omega_n^2 + \e_{\bk}^2+|\Delta_{\bk}|^2},\nonumber\\
\Delta^d&=&-U^d\int_K \frac{f^2_{\bk}\Delta^d}{\omega_n^2 + \e_{\bk}^2+|\Delta_{\bk}|^2}.
\eea
Since we work in the FL regime, we implement $\int_{\bk}=\nu_{2D}\int\frac{d\theta}{2\pi} d\e$ which leads to the definition of two dimensionless parameters $u_s\equiv \nu_{2D}U^s$ and $u_d\equiv \nu_{2D}U^d$ where $\nu_{2D}$ is the 2D density of states at the fermi surface. This model is then easily solved (see Appendix A) and the resulting phase diagram is schematically plotted in Fig. \ref{fig:1}a.

The quantities that change with $u_d$ are $T_c,~\Delta^s,~$and $\Delta^d$. To remove the energy cut-off ($\Lambda$) dependence of our results, these quantities are normalized to $\Delta^s_0$, the gap value at $u_d=0$ (the pure $s-$wave state). We thus work with the normalized parameters:\\
$\alpha_s\equiv \Delta^s/\Delta^s_0,~\alpha_d\equiv \Delta^d/\Delta^s_0,~\eta\equiv \alpha_d/\alpha_s$.\\
The most relevant points that define the boundaries of the $s+id$ phase are at $T=T_c$ and $T=0$. $T_c$ (obtained by setting $\alpha_{s,d}\rightarrow 0$) across the phase diagram is given by \bea\label{eq:tc1}
\text{ln}\frac{2\gamma\Lambda}{\pi T_c}=\text{min}\left\{-\frac{1}{u_s},-\frac{1}{u_d}\right\}>0,
\eea
where $\gamma\approx1.78$. The boundaries of the $s+id$ state can be found after rewriting the self consistency equations as (see Appendix A for details)
\bea\label{eq:10}
\text{ln}\alpha_s&=& -\int\frac{d\theta}{2\pi}\text{ln}\sqrt{1+ \eta^2f^2_{\bk}},\nonumber\\
\frac{1}{u_s}-\frac{1}{u_d}&=& -\int\frac{d\theta}{2\pi}(f^2_{\bk}-1)\text{ln}\sqrt{1 + \eta^2f^2_{\bk}}
\eea
and setting $\eta\rightarrow 0$ and $\eta\rightarrow\infty$ in the second equation. This results in the $s$/$s+id$ boundary at $u_s=u_d$ and the $s+id$/$d$ boundary at $u_d = 2u_s/(2+u_s)$.\cite{Chubukov_sid} Fig. \ref{fig:1}b shows the calculated gaps $\alpha_{s,d}$ as $u_d$ is tuned through the $s+id$ state.

\subsection{Collective modes-formulation and results}
We employ standard linear response to study the collective modes in this system. We provide a simple derivation for the 1-band model as this will help us generalize the formula to the multi-band case with ease. The Hamiltonian in the SC state is given by
\bea\label{eq:Ham}
H&=&\sum_k\Psi^{\dag}_{\bk}\mathcal{H}_{\bk}\Psi_{\bk}, ~~\text{where}~~\Psi^{\dag}_{\bk}=\left(c^{\dag}_{\bk\up}, c_{\bk\down}\right),\nonumber\\
\mathcal{H}_{\bk} &=& \e_{\bk}\sigma_3 - \Delta^R_{\bk}\sigma_1 + \Delta^I_{\bk}\sigma_2,\nonumber\\
\Delta^*_{\bk} &=& -\sum_{\bq}V(\bk,\bq)\langle  c^{\dag}_{-\bq\up} c^{\dag}_{\bq\down} \rangle,
\eea
where $\sigma_0$ is a $2\times2$ identity matrix, $\vec{\sigma}$ is a vector of Pauli matrices and $R,~I$ stand for real and imaginary part of the order parameter. The perturbing fields(originating from fluctuations of density and the order parameter) that couple to this Hamiltonian have the form:
\bea\label{eq:Per}
\delta\mathcal{H}_{\bq,\bk}(t) &=& \left(\delta D_{\bq,\bk}\sigma_3 - \delta\Delta^R_{\bq,\bk}\sigma_1 + \delta\Delta^I_{\bq,\bk}\sigma_2\right)e^{-i\omega t}.
\eea
It is convenient to deal with perturbing fields independent of the internal variable $\bk$ in order to formulate the linear response problem. This is achieved by writing
\bea\label{eq:int}
\delta\Delta^j_{\bq,\bk} &=& \sum_L\delta\Delta^{j,L}_{\bq}f^L_{\bk},\nonumber\\
\delta D_{\bq,\bk} &=& \sum_L\delta D_{\bq}^Lf^L_{\bk},
\eea
where $L$ are the different orthogonal angular momentum channels. (These will correspond to the irreducible representations of the point symmetry group in the presence of a lattice). In the usual case of density fluctuations, there is no dependence of $\delta D_{\bq,\bk}$ on $\bk$ (as it corresponds to non-equilibrium fluctuations of total density). This is why the higher angular momentum channels, in this simple model, are not affected by the Coulomb force. This will be utilized later. For now, proceeding in complete generality, we are then led to
\bea\label{eq:int2}
&\delta H_{\bq}(t)&\nonumber\\
&=&\sum_{k,L}f^L_{\bk}\Psi^{\dag}_{\bk}\left(\delta D^L_{\bq}\sigma_3  -\delta\Delta^{R,L}_{\bq}\sigma_1 +\delta\Delta^{I,L}_{\bq}\sigma_2 \right)\Psi_{\bk}e^{-i\omega t},\nonumber\\
&=&\sum_{L,i}\delta F^L_i(\bq)\mathcal{R}^L_i(\bq)e^{-i\omega t}.
\eea
where
\bea\label{eq:int3}
\mathcal{R}^L_i(\bq)&=&\sum_k f^L_{\bk}\Psi^{\dag}_{\bk}\sigma_i\Psi_{\bk},
\eea
and the perturbing field
\beq\label{eq:pert}
\delta F^L_i = (-\delta\Delta^{R,L},\delta\Delta^{I,L}, \delta D^L).
\eeq
The self-consistency equation in Eq. (\ref{eq:Ham}) is then written as:
\bea\label{eq:decoup}
(\Delta^*)^L &=& -\sum_{\bk}V^{LL'}f^{L'}_{\bk}\langle  c^{\dag}_{-\bk\up} c^{\dag}_{\bk\down} \rangle,
\eea
where $V^{LL'}$ is defined through
\bea\label{eq:decoup}
V(\bk,\bq) &=& \sum_{L,L'}V^{LL'}f^{L}_{\bk}f^{L'}_{\bq}.
\eea
Starting from
\bea\label{eq:int4}
\delta\Delta^*_{\bq} &=& -\sum_{\bk}V(\bq,\bk)\delta\langle  c^{\dag}_{-\bk\up} c^{\dag}_{\bk\down} \rangle,
\eea
we can make use of the following relations
\bea\label{eq:not}
\Psi^{\dag}\sigma_1\Psi &=& c^{\dag}_{\up}c^{\dag}_{\down}-c_{\up}c_{\down},\nonumber\\
\Psi^{\dag}\sigma_2\Psi &=& -i\left(c^{\dag}_{\up}c^{\dag}_{\down}+c_{\up}c_{\down}\right),\nonumber\\
\Psi^{\dag}\sigma_3\Psi &=& c^{\dag}_{\up}c_{\up}+c^{\dag}_{\down}c_{\down},
\eea
to write down,
\bea\label{eq:int5}
2\delta\Delta^R_{\bq} &=& -\sum_{\bk}V(\bq,\bk)\delta\langle  \Psi^{\dag}_{\bk}\sigma_1\Psi_{\bk} \rangle,\nonumber\\
-2\delta\Delta^I_{\bq} &=& -\sum_{\bk}V(\bq,\bk)\delta\langle  \Psi^{\dag}_{\bk}\sigma_2\Psi_{\bk} \rangle,\nonumber\\
\delta D_{\bq} &=& V_{\bq}\sum_{\bk}\delta\langle  \Psi^{\dag}_{\bk}\sigma_3\Psi_{\bk} \rangle,\nonumber\\
\eea
where $V_{\bq}=2\pi e^2/q$. Following the argument around Eq. (\ref{eq:decoup}) and abbreviating $V^{LL}$ as $V^L$, we can write Eq. (\ref{eq:int5}) as:
\bea\label{eq:int8}
2\delta\Delta^{R,L}_{\bq} &=& -V^{L}\delta\langle\mathcal{O}^{L}_1\rangle,\nonumber\\
-2\delta\Delta^{I,L}_{\bq} &=& -V^{L}\delta\langle\mathcal{O}^{L}_2\rangle,\nonumber\\
\delta D^L_{\bq} &=& V^{L}_{\bq}\delta\langle\mathcal{O}^{L}_3\rangle.\nonumber\\
\eea
We will work in the limit of $\bq\rightarrow 0$ so that $\bq$ will only be retained in $D_{\bq}$ to account for the singular nature of the Coulomb interaction.

The statement of linear response is that the change in an operator due to the applied perturbation is given by (recalling that $\mathcal{G}=-\langle\Psi\Psi^{\dag}\rangle$)
\bea\label{eq:Lin_Response}
\delta\langle\mathcal{R}_i^L\rangle(Q) = \sum_{j,L'}\Pi^{LL'}_{ij}(Q)\delta F^{L'}_j(Q),\nonumber\\
\Pi^{LL'}_{ij}(Q) = \int_K f^L_{\bk}f^{L'}_{\bk}\text{Tr}\left[G(K)\sigma_iG(K+Q)\sigma_j\right].
\eea
Combining Eqs. (\ref{eq:int4})-(\ref{eq:Lin_Response}), we arrive at the one-band system of equations:
\bea\label{eq:1band}
\sum_{j,L'}\left\{\Pi_{ij}^{LL'}-2[V^L]^{-1}\delta^{LL'}\delta^{ij}\right\}\delta F_j^{L'}&=&0.\nonumber\\
\eea
See Appendix B for explicit form of the mode equation. The non-trivial solutions of this set of equations are the collective modes of the system.

\subsubsection{General considerations}
We see from Eq. (\ref{eq:1band}), that the collective modes can be found once the interactions $V^L$ and the polarization bubbles $\Pi^{LL'}_{ij}$ are known. These depend on the specifics of a microscopic model and can be easily computed. In a particle-hole symmetric system, quite generally, we will have:
\bea\label{eq:pis}
\Pi^{LL'}_{11}(i\Omega_n)&=&\int_K\frac{2f^L_{\theta}f^{L'}_{\theta}}{D_+D_-}\left[-\omega_+\omega_--\e^2+\Delta^2_R-\Delta^2_I\right],\nonumber\\
\Pi^{LL'}_{22}(i\Omega_n)&=&\int_K\frac{2f^L_{\theta}f^{L'}_{\theta}}{D_+D_-}\left[-\omega_+\omega_--\e^2-\Delta^2_R+\Delta^2_I\right],\nonumber\\
\Pi^{LL'}_{33}(i\Omega_n)&=&\int_K\frac{2f^L_{\theta}f^{L'}_{\theta}}{D_+D_-}\left[-\omega_+\omega_-+\e^2-\Delta^2_R-\Delta^2_I\right],\nonumber\\
\Pi^{LL'}_{13}(i\Omega_n)&=&\int_K\frac{2f^L_{\theta}f^{L'}_{\theta}}{D_+D_-}\left[\Omega_n\Delta_I\right]~=~-\Pi^{LL'}_{31}(i\Omega_n),\nonumber\\
\Pi^{LL'}_{23}(i\Omega_n)&=&\int_K\frac{2f^L_{\theta}f^{L'}_{\theta}}{D_+D_-}\left[-\Omega_n\Delta_R\right]~=~-\Pi^{LL'}_{32}(i\Omega_n),\nonumber\\
\Pi^{LL'}_{12}(i\Omega_n)&=&\int_K\frac{2f^L_{\theta}f^{L'}_{\theta}}{D_+D_-}\left[2\Delta_R\Delta_I\right]~=~\Pi^{LL'}_{21}(i\Omega_n).\nonumber\\
\eea
$D_{\pm} = \left(\omega_m\pm\frac{\Omega_n}{2}\right)^2 + \e^2 + |\Delta|^2$. $\Delta_{R,I}$ are the ground state properties and are taken as input from the analysis in the previous section. Although the temperature evolution can be tracked, we shall perform calculations at $T=0$ as the calculations are tractable and already very informative. We make use of the following integrals:
\bea\label{eq:e}
\int_K\frac{\omega_m^2}{D_+D_-}&=&\frac{1}{\Delta_s^0}\int_{\bk}\frac{1}{4E},\\
\int_K\frac{1}{D_+D_-}&=&\frac{1}{(\Delta_s^0)^3}\int_{\bk}\frac{1}{4E}\frac{1}{E^2 + \left(\frac{\Omega_n}{2\Delta^0_s}\right)^2},
\eea
where $E^2=\left(\frac{\e_{\bk}}{\Delta_s^0}\right)^2 + \alpha_s^2 + \alpha_d^2f^2_{\bk}$; and
\bea\label{eq:e2}
\nu_{2D}\Delta^0_sI_0(i\Omega_n)=\int_{\bk}\frac1E\frac{1}{E^2 + \left(\frac{\Omega_n}{2\Delta_s^0}\right)^2},\\
\nu_{2D}\Delta^0_sI_2(i\Omega_n)=\int_{\bk}\frac1E\frac{f^2_{\theta}}{E^2 + \left(\frac{\Omega_n}{2\Delta_s^0}\right)^2},\\
\nu_{2D}\Delta^0_sI_4(i\Omega_n)=\int_{\bk}\frac1E\frac{f^4_{\theta}}{E^2 + \left(\frac{\Omega_n}{2\Delta_s^0}\right)^2}.
\eea
where $f^s_{\theta}=1;~f^d_{\theta}=\sqrt{2}\cos2\theta$. It is worth noting that in the $s-$wave ground state ($\alpha_d=0$), $I_2=I_0$ and $I_4=\frac32I_0$. Analytic continuation to real frequencies is performed by $i\Omega_n \rightarrow \Omega + i\delta$. As one tunes $u_d$, the ground state changes, and the $\Pi^{LL'}_{ij}$'s need to be calculated at every $u_d$.

In what follows, we will ignore the coupling of the collective modes to the charge sector as justified in the Introduction. We refer the reader to discussion in Ref. \onlinecite{SM_AVC} and references therein where explicit coupling to the charge sector, within the same formalism, is presented. The results from now on therefore pertain, strictly speaking, to the collective modes in a `neutral' SC. This simplification allows the formulation of the whole problem in a $4\times4$ space of $\delta\Delta_{s,d}^{R,I}$. In the chosen gauge (where the $s-$wave condensate is chosen to be real), $R$ maps onto the amplitude sector and $I$ maps onto the phase sector. We now investigate the individual cases.

\subsubsection{Collective modes - $s$ wave ground state}

It can be seen from Eq. (\ref{eq:pis}) that in a pure angular momentum ground state (pure $s$ or pure $d$), $\Pi^{LL'}=0$ if $L\neq L'$ (due to orthogonality) and in non-complex order parameter ground state, $\Pi_{12}=0$. Thus, in the $s$ wave ground state, the only surviving bubbles are:
\bea\label{eq:swave}
\Pi^{ss}_{11}(\Omega)&=& \nu_{2D}\left[-4L_g + \left(1-\left(\frac{\Omega}{2\Delta_0^s}\right)^2\right)I_0(\Omega)\right],\nonumber\\
\Pi^{ss}_{22}(\Omega)&=& \nu_{2D}\left[-4L_g  -\left(\frac{\Omega}{2\Delta_0^s}\right)^2I_0(\Omega)\right],\nonumber\\
\Pi^{dd}_{ii}(\Omega)&=& I_0\rightarrow I_2~(=I_0),~i\in(1,2),
\eea
where $L_g=\int_{\bk}\frac{1}{4E}=\frac12 \text{ln}\frac{2\Lambda}{\Delta^s_0}$. This implies that (1) $s$ and $d$ channels are completely decoupled and (2) amplitude and phase sector are completely decoupled. As a result, the collective mode equation (det[Eq. (\ref{eq:1band})$=0$]) in the amplitude(phase) sector of angular momentum $L$ reads
\beq\label{eq:amps}
\Pi^{LL}_{11(22)}-\frac{2}{V^L} = 0,
\eeq
where, $V^L$ is the interaction in the $L^{th}$ angular momentum channel ($s$ or $d$).

\emph{Collective modes in the amplitude sector:} The solutions to the mode equation are contained in
\bea\label{eq:ssamp1}
&&\left(1-\left(\frac{\Omega}{2\Delta_0^s}\right)^2\right)I_0(\Omega) = 0,\\
\label{eq:ssamp2}
&&\left(1-\left(\frac{\Omega}{2\Delta_0^s}\right)^2\right)I_2(\Omega) = \frac{2}{u_d}-\frac{2}{u_s}.
\eea
Neither of these equations has an undamped solution (see Appendix C).

\emph{Collective modes in the phase sector:} The solutions to the mode equation are contained in
\bea\label{eq:ssphase1}
\left(\frac{\Omega}{2\Delta_0^s}\right)^2I_0(\Omega) = 0,\\
\label{eq:ssphase2}
-\left(\frac{\Omega}{2\Delta_0^s}\right)^2I_2(\Omega) = \frac{2}{u_d}-\frac{2}{u_s}.
\eea
The first equation yields the soft BAG mode ($\Omega=0$) and the second (arising from the $d-$wave sector) yields the well known BS mode.\cite{BardasisSchrieffer61,zawadowskii} Looking Eq. (\ref{eq:tc1}) and Eq. (\ref{eq:ssphase2}) we see that the BS mode frequency is related to the competing $T_c$ values for the $s$ and $d$ channels. Thus detecting the BS mode at low temperatures and recording the $T_c$ of a sample gives direct quantitative estimate of the competing $d$-wave pairing interaction.\cite{DevereauxScalapino09} This feature will change for a multiband system.

The above modes lie below the minimum gap in the system and hence are not damped. Notice that at boundary of the $s$ and $s+id$ state, where $u_s=u_d$, the BS mode softens as expected.

\subsubsection{Collective Modes in the $s+id$ ground state}
The collective modes in the $s+id$ ground state are interesting. The surviving bubbles in this state are:
{\small
\bea\label{eq:e18}
\Pi^{ss}_{11}(\Omega)&=&\nu_{2D}\left[-2\int_{\theta}\ln\frac{2\Lambda}{|\Delta_{\theta}|} + \left(\alpha_s^2-\left(\frac{\Omega}{2\Delta_0^s}\right)^2\right)I_0(\Omega)\right],\nonumber\\
\Pi^{ss}_{22}(\Omega)&=& \nu_{2D}\left[-2\int_{\theta}\ln\frac{2\Lambda}{|\Delta_{\theta}|} + \alpha_d^2I_2(\Omega)-\left(\frac{\Omega}{2\Delta_0^s}\right)^2I_0(\Omega)\right],\nonumber\\
\Pi^{dd}_{11}(\Omega)&=&\nu_{2D}\left[-2\int_{\theta}f^2_{\theta}\ln\frac{2\Lambda}{|\Delta_{\theta}|} + \left(\alpha_s^2-\left(\frac{\Omega}{2\Delta_0^s}\right)^2\right)I_2(\Omega)\right],\nonumber\\
\Pi^{dd}_{22}(\Omega)&=& \nu_{2D}\left[-2\int_{\theta}f^2_{\theta}\ln\frac{2\Lambda}{|\Delta_{\theta}|} + \alpha_d^2I_4(\Omega)-\left(\frac{\Omega}{2\Delta_0^s}\right)^2I_2(\Omega)\right],\nonumber\\
\Pi^{sd}_{12}(\Omega)&=& \nu_{2D}\alpha_s\alpha_dI_2(\Omega),
\eea
}
where $\alpha_{s,d}$ are to be found from $T=0$ solutions of the gap equation as discussed in Appendix A. The non-zero $\Pi^{sd}_{12}$ couples the $s$ and $d$ channels and also the amplitude and phase sector. This is expected from Eq. (\ref{eq:pis}) because the ground state itself is a mixture of the two angular momentum channels and the order parameter is complex. The physical consequence of this non-zero bubble is that the 4 decoupled sectors now coalesce into two $2\times2$ sectors formed out of the\\ (1) $s$-phase and the $d-$amplitude components which yields
\bea\label{eq:ds1}
\left(\frac{\Omega}{2\Delta_0^s}\right)^2I_2\left[-\left(\frac{\Omega}{2\Delta_0^s}\right)^2I_0 + \alpha_s^2I_0 +\alpha^2_dI_2\right]&=&0.
\eea
This sector contains the gauge-mode at $\Omega=0$. \\(2) $s$-amplitude and the $d-$phase components which yields
\bea\label{eq:ds2}
\left(\alpha_d^2I_4 - \left(\frac{\Omega}{2\Delta_0^s}\right)^2I_2\right)\left(\alpha_s^2-\left(\frac{\Omega}{2\Delta_0^s}\right)^2\right)I_0-\alpha_s^2\alpha_d^2I_2^2&=&0.\nonumber\\
\eea
This contains the collective  mode with mixed symmetry that adiabatically continues to the Bardasis-Schrieffer mode in the $s$-wave phase; we will refer to this mode henceforth as the mixed-symmetry Bardasis-Schrieffer mode (MSBS). This mixed symmetry mode is the analog of the $d+id'$ and $p+ip'$ clapping modes discussed in Refs. \onlinecite{VW,Hirschfeld92,Balat}.

The $d-$ ground state can be similarly worked out (see Appendix C). There are no collective modes (other than the BAG mode) that propagate and hence we do not dwell on this further.

Fig. \ref{fig:1}(c) traces all the collective modes below the minimum gap across the phase diagram. It is worth noting that all the collective modes could be found essentially from one mode equation.

\section{Collective modes in a 3-pocket model}\label{sec:3poc}
We now move to the 3-pocket model which is more relevant for the FeSCs. Other than the multiband aspect, the approach is identical to the 1 band model. We shall thus focus on discussing the results and highlight the differences with the 1-band model.
\subsection{Model and phase diagram}
\begin{figure}[htp]
\includegraphics[width=.99\columnwidth]{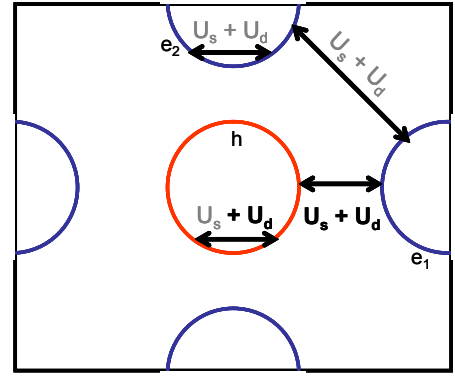}
\caption{\label{fig:scceem} The interactions in a 3 pocket (1 hole and 2 electron) model. The interactions in light grey font are unimportant as far as the main message of the work with applications to hole doped FeSC is concerned and are thus set to zero.}
\end{figure}

Here we study a prototypical FeSC system with one $\Gamma$ centered hole pocket and two $M$ centered electron pockets (the latter two are from the same band-and hence are related by symmetry). The model and the pairing interactions between the fermions is schematically shown in Fig. \ref{fig:scceem}. The interactions in explicit form can be written as (only the leading harmonics are retained; $e_1\rightarrow +1$ and $e_2\rightarrow-1$):
\bea\label{eq:intform}
V_{hh}(\bk,\bp)&=& U^s_{h} + U^d_{h}f_kf_p,\nonumber\\
V_{e_1e_1}(\bk,\bp)&=& U^s_{e} + U^d_{e}(1)(1),\nonumber\\
V_{e_2e_2}(\bk,\bp)&=& U^s_{e} + U^d_{e}(-1)(-1),\nonumber\\
V_{e_1e_2}(\bk,\bp)&=& U^s_{e_1e_2} + U^d_{e_1e_2}(1)(-1),\nonumber\\
V_{he_1}(\bk,\bp)&=& U^s_{he} + U^d_{he}f_k(1),\nonumber\\
V_{he_2}(\bk,\bp)&=& U^s_{he} + U^d_{he}f_k(-1),\nonumber\\
\eea
where $f_k = \sqrt{2}\cos\theta_k$. The self consistency gap equations for this model read:
\bea\label{eq:gap eq}
\Delta^h_{\bp}&=& -\int_{\bk}\left[V^h_{\bp,\bk}\Delta^h_{\bk}W^h_{\bk}+
V^{he_1}_{\bp\bk}\Delta^{e_1}_{\bk}W^{e_1}_{\bk} + ~(e_1\leftrightarrow e_2)\right],\nonumber\\
\Delta^{e_1} &=& -\int_{\bk}\left[V^{he_1}_{\bp\bk}\Delta^{h}_{\bk}W^{h}_{\bk}\right],\nonumber\\
\Delta^{e_2} &=& -\int_{\bk}\left[V^{he_2}_{\bp\bk}\Delta^{h}_{\bk}W^{h}_{\bk}\right],\nonumber\\
\eea
where $W^x_{\bk}=\frac{1}{2E^x_{\bk}}\tanh\frac{E^x_{\bk}}{2T}$, with $x\in\{h,e_1,e_2\}$. We further assume, for the sake of simplicity of presentation, that the electron and hole bands have identical dispersions: resulting in the same density of states. This choice of interactions requires the gap structure to assume the form
\bea\label{eq:char}
\Delta^h_{\bp}&=&\Delta^h_s + \Delta^h_d f_{\bp},\nonumber\\
\Delta^{e_1}_{\bp}&=&\Delta^{e}_s + \Delta^{e}_d,\nonumber\\
\Delta^{e_2}_{\bp}&=&\Delta^{e}_s - \Delta^{e}_d.
\eea

\begin{figure*}[htp]
$\begin{array}{ccc}
\includegraphics[width=0.6\columnwidth]{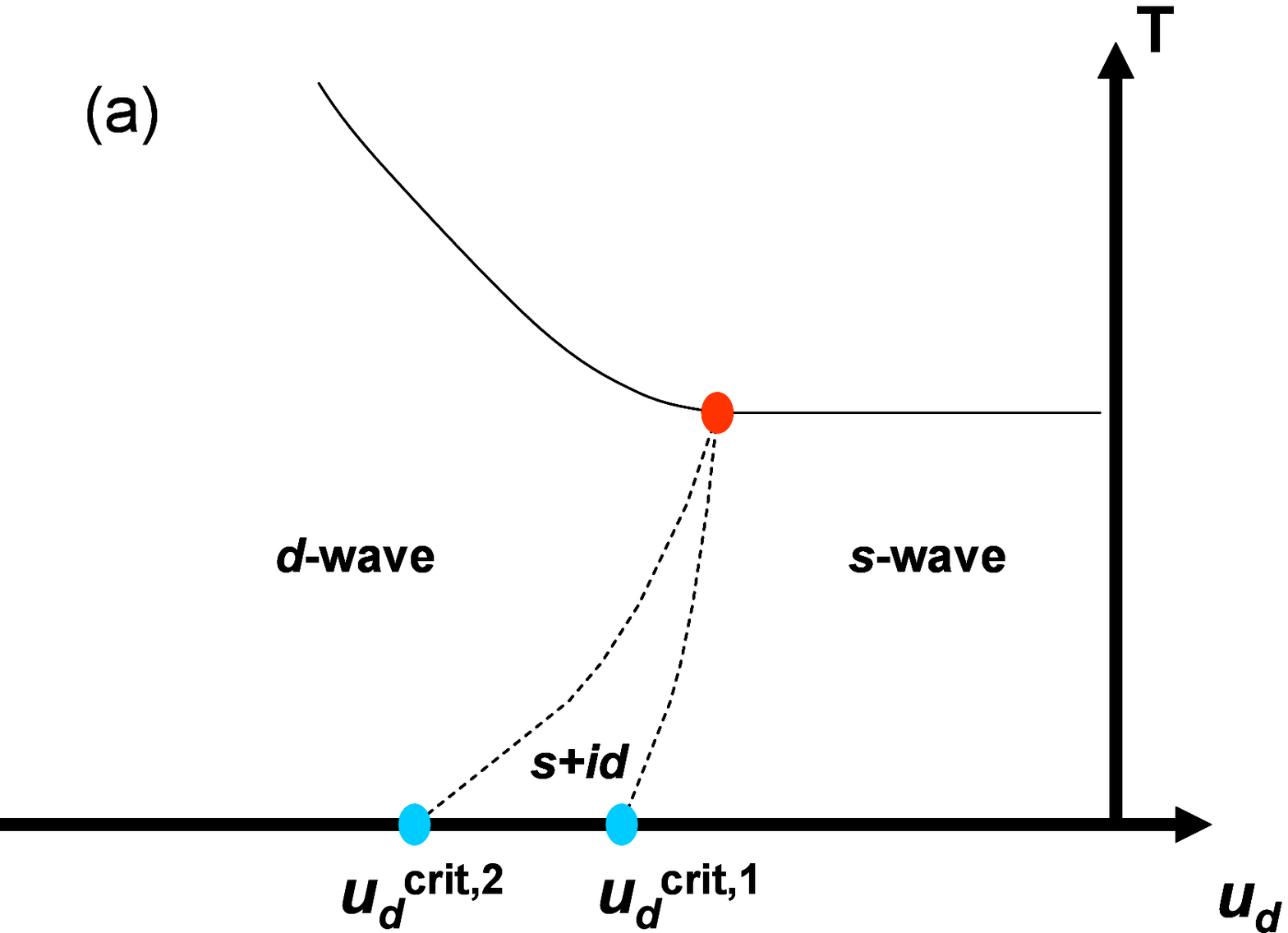}&
\includegraphics[width=0.7\columnwidth]{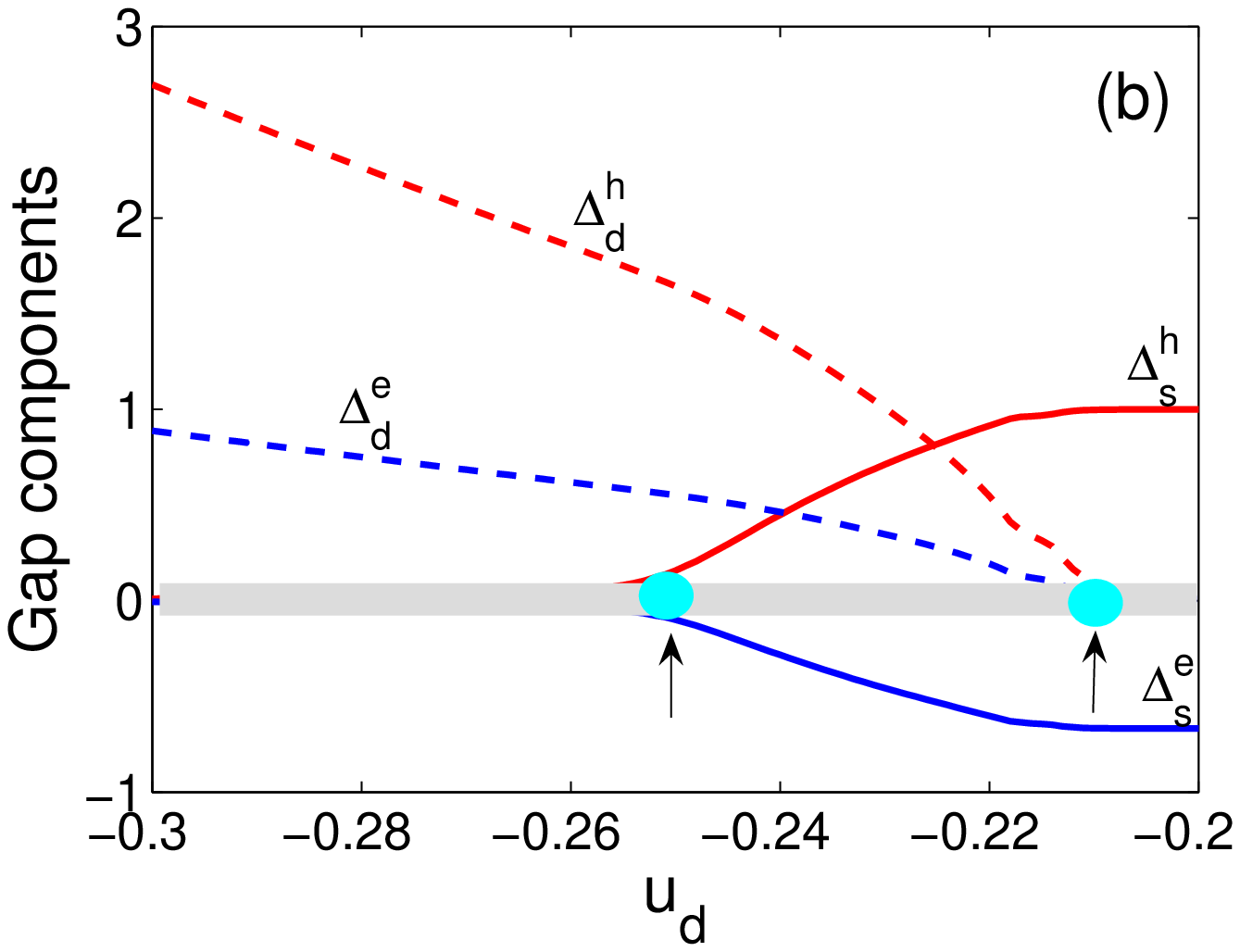}&
\includegraphics[width=0.7\columnwidth]{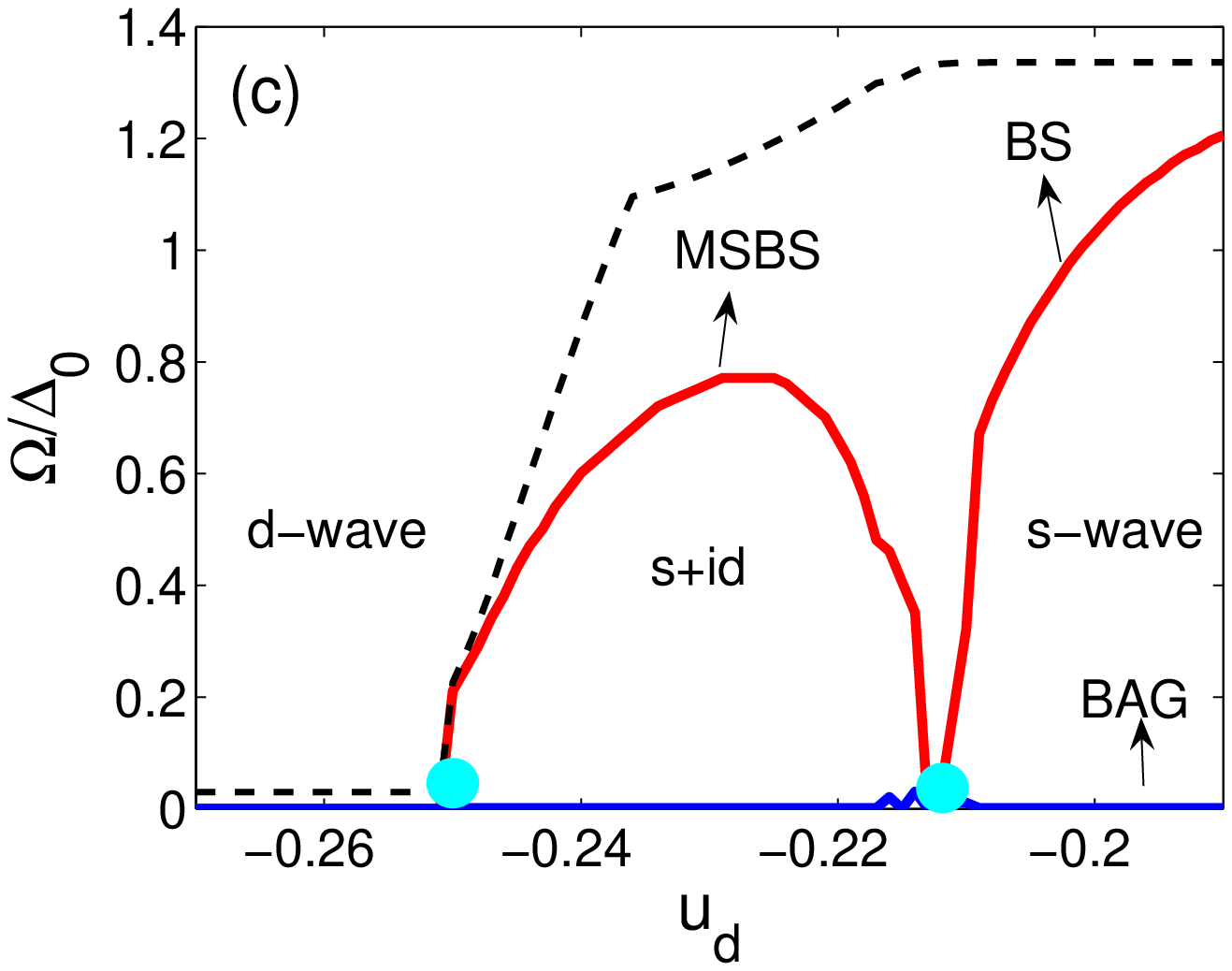}
\end{array}$
\caption{\label{fig:2band phase} (a) The phase diagram in the 3-pocket model. (b) The evolution of the gap components with $u_d$ in the 3-pocket model at $T=0$. The boundaries of the $s+id$ phase are marked with arrows. Red(Blue) represents the hole(electron) pocket and the solid(dashed) line corresponds to  $s-$wave($d-$wave) component of the gap. The grey line at zero is the size of error in the numerical calculation due to the choice of grid and resolution parameters. (c): The (undamped) collective modes across the phase diagram in  different channels. The dashed black line is $2\Delta_{\text{min}}$ in the system. The dots indicate the $s+id$ phase boundaries at $T=0$. In all the figures $z = 1/2$ and $v_s=0.2$.}
\end{figure*}

To minimize the parameter space we set the following interactions to zero: $U^s_h$, $U^{s,d}_{e,e_1e_2}$. The rationale behind this lies in the fact that we want to model a system driven to a $d-$wave state by the hole pockets (hence we retain $U^d_h$) and the $d-$wave character in the rest is induced due to the interband coupling terms (hence we retain $U^d_{he}$). The reason behind the choice of this model is related to our desire to eventually address the Raman experiments on the hole doped FeSC, and is explained further in Sec. \ref{sec:disc}. The $s-$wave character in this system driven by $U^s_{he}$. For brevity, we introduce the following dimensionless constants for the interactions
\beq\label{eq:ddd}
v_{s,d}\equiv \nu_{2D}U^{s,d}_{he},~u_d\equiv \nu_{2D}U^d_{h}.
\eeq
Recognizing the $v_s$ and $u_d$ are the main ingredients for our problem and that $v_d$ is only needed to induce SC in the electron pockets, we set $v_d=z u_d$; where the ratio $z$ is set to some constant. Keeping $v_s$ fixed (as in the 1-band case), the system now has $u_d$ (and $z$ if one so desires) as the tuning parameter of the model. This is sufficient to generate the $s$ to $s+id$ to $d$ phase diagram. We will then need to define a few more dimensionless parameters in analogy with the 1-band case
\bea\label{eq:param}
&&\alpha^x_s\equiv\Delta^x_s/\Delta_0;~~\alpha^x_d\equiv\Delta^x_d/\Delta_0,~x\in(h,e),\nonumber\\
&&r_s\equiv\frac{\alpha^e_s}{\alpha^h_s};~~r_d\equiv\frac{\alpha^e_d}{\alpha^h_d},
\eea
where $\Delta_0$ is the $s$-wave gap on the hole pocket when $v_d=0$. This model is easily solved at $T=0$ and $T=T_c$ (see Appendix D). The results for the boundaries of the $s+id$ state are given as follows: at $T=T_c$, the critical $u_d(<0)$ is the solution to (larger $|u_d|$ favors a $d-$wave state)
\beq\label{eq:deg2}
\left(\frac{zu_d}{v_s}\right)^2= 1 + \frac{1}{\sqrt{2}}\left(\frac{u_d}{v_s}\right).
\eeq
There are two points marking the boundary of the $s+id$ state at $T=0$. The $s$-wave side boundary (setting $\Delta^{e,h}_d\rightarrow 0$ and requiring $r_d$ to be arbitrary) yields
\bea\label{eq:T0t2}
&&\frac{1}{r_s}-2r_s=2v_s\ln|r_s|, \nonumber\\
&&u_d^{\text{crit},1} = \frac{-r_s-\sqrt{r_s^2+4z^2}}{2z^2}v_s.
\eea
The $d-$wave side boundary (with $\Delta^{h,s}_{s}\rightarrow 0$, $r_s$ arbitrary) $u_d^{\text{crit},2}$ is the solution to
\bea\label{eq:T05}
\frac{1}{2v_s^2}&=&\left[-\frac{r_d}{zu_d}+\frac12\right]\left[-\frac{r_d}{zu_d}+c_2-\ln|r_d|\right],\nonumber\\
\label{eq:TOF3}
\text{and $r_d$ satisfies}&&\nonumber\\
\frac{1}{r_d}-2r_d&=&zu_d\left(c_2-\ln|r_d|\right),
\eea
where $c_2\equiv\int f^2\ln |f| =0.153$. These boundaries are shown in Fig. \ref{fig:2band phase}(a). The detailed solution for the gap components as a function of $u_d$ presented in Fig. \ref{fig:2band phase}(b).

\subsection{Collective modes}
For multiband systems, we follow the same procedure to derive the collective mode equation. Unless we consider Cooper pairing between different bands, there will be no interband $\Pi^{LL'}_{ij}$'s of the type $\int G^hG^e$. Then, Eq. (\ref{eq:1band}) will be generalized to matrix equation in the band space. $\Pi^{LL'}_{ij}$ is then evaluated in the relevant ground state. This generalization is given by:
\bea\label{eq:Nband}
\sum_{j,L',b}\left\{\Pi_{ij,a}^{LL'}[\delta_{ab}]-2[V_L^{-1}]_{ab}\delta^{LL'}\delta^{ij}\right\}\delta F_{j,b}^{L'}&=&0,
\eea
where $a,~b$ are the band indices. The explicit form of this equation is discussed in  Appendix E. The only off-diagonal elements (in the band space) arise from $[V^L]^{-1}$. As before, we drop the coupling to the density channel and work with a $2_{s/d}\times2_{\text{real/imag}}\times2_{\text{bands}}= 8\times8$ matrix space. The interaction matrices in each angular momentum channel are given by
\beq
[V_s]=
\left(\begin{array}{cc}
0&v_s\\
v_s&0\\
\end{array}
\right),~[V_d]=
\left(\begin{array}{cc}
u_d&v_d\\
v_d&0
\end{array}
\right).
\eeq

\subsubsection{Collective modes in the $s$-wave ground state}
As in the 1-band scenario, $\Pi^{sd}$ and $\Pi_{12}$ are zero due to the symmetry of the ground state. The non-zero $\Pi$'s are given by
\bea\label{eq:spi}
\Pi^{ss}_{11,h}&=&\nu_{2D}\left[-4L_h + \left(1-\left(\frac{\Omega}{2\Delta_0}\right)^2\right)I^h_0(\Omega)\right],\nonumber\\
\Pi^{ss}_{11,e}&=&\nu_{2D}\left[-8L_e + 2\left((\alpha^e_s)^2-\left(\frac{\Omega}{2\Delta_0}\right)^2\right)I^e_0(\Omega)\right],\nonumber\\
\Pi^{ss}_{22,h}&=&\nu_{2D}\left[-4L_h -\left(\frac{\Omega}{2\Delta_0}\right)^2I^h_0(\Omega)\right],\nonumber\\
\Pi^{ss}_{22,e}&=&\nu_{2D}\left[-8L_h -2\left(\frac{\Omega}{2\Delta_0}\right)^2I^e_0(\Omega)\right],\nonumber\\
\eea
where $L_{h,e}=\frac12 \ln\frac{2\Lambda}{|\Delta^{h,e}_s|}$. $\Pi^{dd}$ involves changing $I_0\rightarrow I_2$, but in the $s-$wave ground state $I_0 = I_2$. Also for $x\in(h,e)$,
\bea\label{eq:somedef}
\nu_{2D}\Delta_0I^x_m(i\Omega_n)&=&\int_{\bk}\frac{1}{E^x}\frac{(f_x)^m}{(E^x)^2+\left(\frac{\Omega_n}{2\Delta_0}\right)^2},\nonumber\\
(E^x)^2&=& \left(\frac{\e^x}{\Delta_0}\right)^2 + (\alpha^x_s)^2 + (\alpha^x_d)^2f^2_x.
\eea

\begin{figure}[htp]
$\begin{array}{c}
\includegraphics[width=0.8\columnwidth]{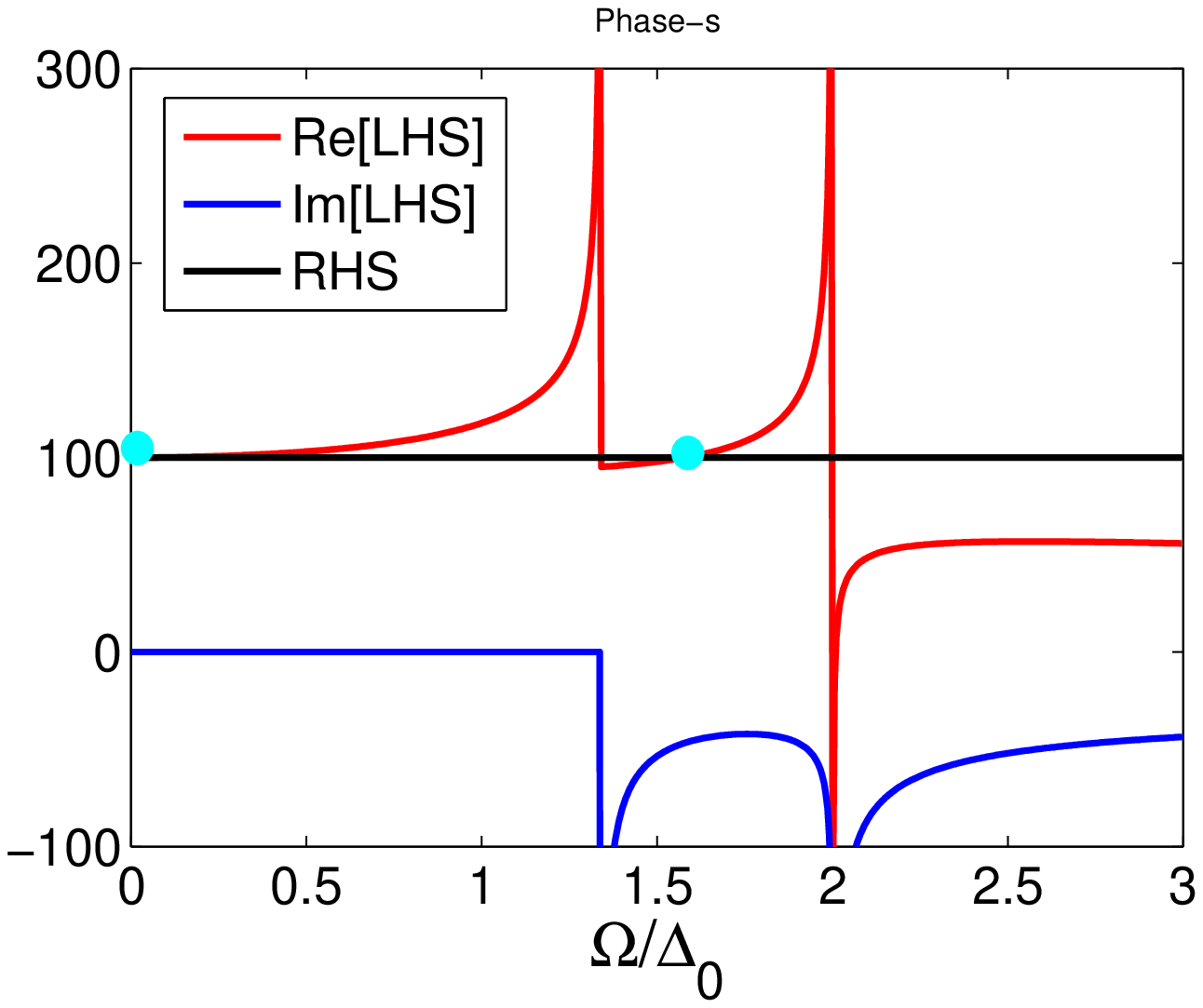}\\
\includegraphics[width=0.8\columnwidth]{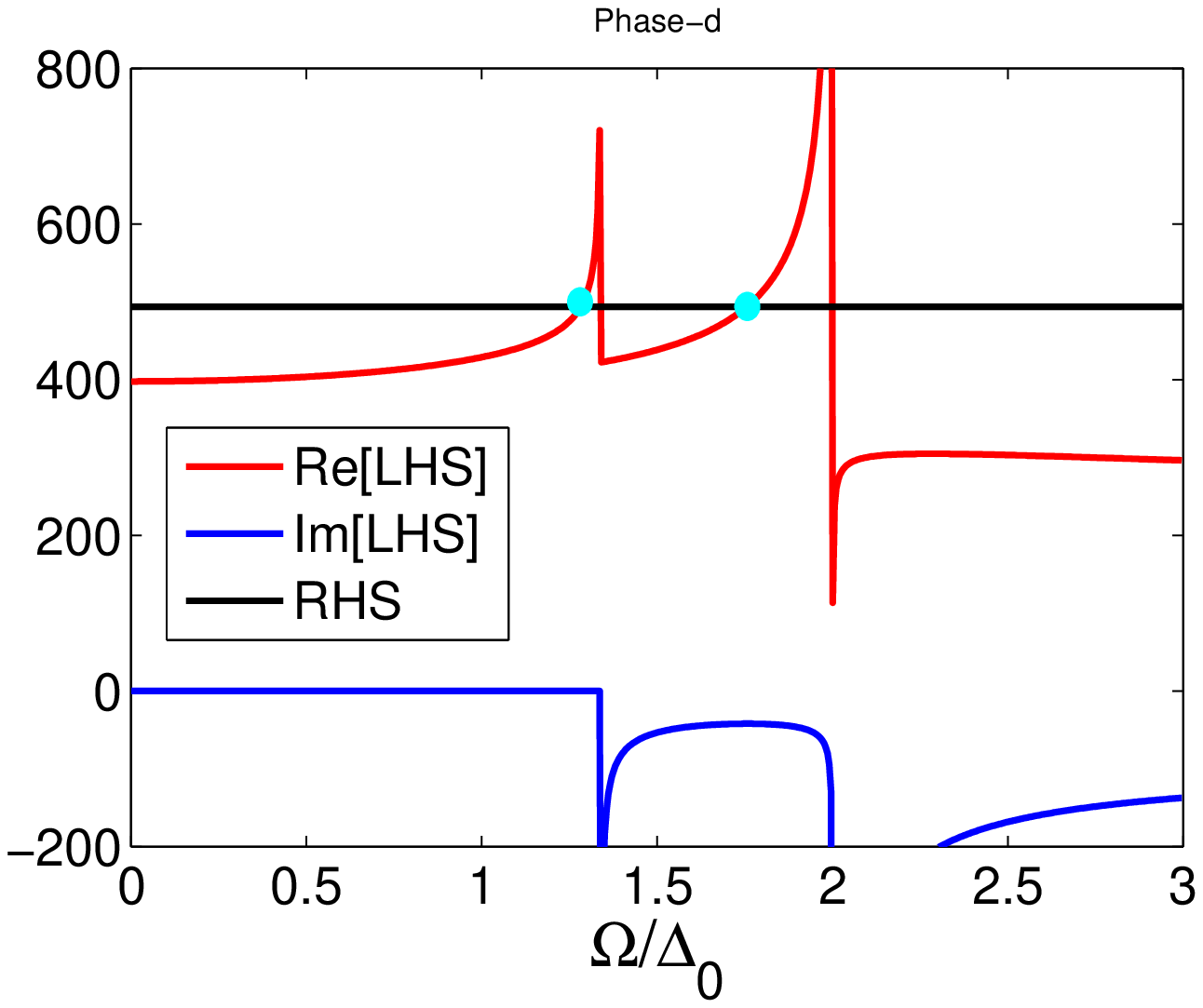}
\end{array}$\caption{
\label{fig:sred} Collective mode solutions in the multi-band $s-$wave ground state. Top: The $s$-wave order parameter phase fluctuation sector -- solution to Eq. (\ref{eq:smodes2}) The blue dots indicate the frequency of the collective mode/resonance(if the imaginary part is non-zero). The $\Omega=0$ solution is the BAG mode; and the $\Omega$ between the coherence peaks is the Leggett resonance (as in MgB$_2$\cite{MGB2}). Bottom: The $d$-wave order parameter phase fluctuation sector -- solution to Eq. (\ref{eq:smodes4}). The dots indicate the frequency of the collective mode/resonance. There is a conventional BS mode below the minimum gap and another resonance in between the two $s-$wave gaps. Here is used parameters $v_s=0.2$, $u_d=-0.18$, $z=1/2$.}
\end{figure}

For the $s-$wave ground state, sectors in the $s$ and $d$ channels decouple. Further, the amplitude and phase sectors decouple. This leads to the following four decoupled equations:
\bea
\label{eq:smodes1}
\Pi_{11,h}^{ss}\Pi_{11,e}^{ss}&=& \frac{4}{v_s^2},\\
\label{eq:smodes2}
\Pi_{22,h}^{ss}\Pi_{22,e}^{ss} &=& \frac{4}{v_s^2},\\
\label{eq:smodes3}
\Pi_{11,h}^{dd}\left(\Pi_{11,e}^{dd}+\frac{2u_d}{v_d^2}\right) &=& \frac{4}{v_d^2},\\
\label{eq:smodes4}
\Pi_{22,h}^{dd}\left(\Pi_{22,e}^{dd}+\frac{2u_d}{v_d^2}\right) &=& \frac{4}{v_d^2}.
\eea
It helps to note that at $T=0$ the self consistency equations tell us that
\bea\label{eq:TOS}
2L_e&=&-\frac{1}{2r_sv_s},\nonumber\\
2L_h&=&-\frac{r_s}{v_s}.
\eea
Eq. (\ref{eq:smodes1}) is the $s-$amplitude sector which has no solution. Eq. (\ref{eq:smodes2}) is the $s-$phase sector. This sector contains the BAG mode and the damped Leggett mode (blue dots in Fig. \ref{fig:sred}(top) where the real part of LHS $=$ RHS). Using $T=0$ relations, Eq. (\ref{eq:smodes2}) gives
\bea\label{eq:wtf2}
\left(\frac{\Omega}{2\Delta_0}\right)^2\left[4L_hI^e_0 + 4L_eI_0^h + \left(\frac{\Omega}{2\Delta_0}\right)^2I_0^hI_0^e\right]&=&0.
\eea
Since $I_0>0$, the only solution is $\Omega=0$ the BAG mode. The other solution, a Leggett resonance, has an imaginary part and is thus damped. Eq. (\ref{eq:smodes3}) is the $d-$amplitude sector which also has no solution. Eq. (\ref{eq:smodes4}) is the $d-$phase sector which contains the BS mode. Note that there is one true mode (below the minimum gap) and a resonance in the continuum as shown by the blue dots in Fig. \ref{fig:sred}(bottom). To see that this true generically in the model, we again substitute for $\Pi$'s and use $T=0$ relations to we find-
\bea\label{eq:wtf2}
\left(\frac{\Omega}{2\Delta_0}\right)^4 I^h_0I^e_0 + \left(\frac{\Omega}{2\Delta_0}\right)^2\mathcal{A} -\frac{2}{v_d^2}\mathcal{S}&=&0,
\eea
where
\bea\label{eq:wtf3}
\mathcal{S}&=&1-r_s\frac{u_d}{v_s}-\frac{v_d^2}{v_s^2},\\
\mathcal{A}&=&-\frac{2}{v_s}\left(\frac{1}{2r_s}+r_s+\frac{u_dv_s}{2v_d^2}\right)>0.
\eea
Comparing Eq. (\ref{eq:deg2}) and Eq. (\ref{eq:wtf3}) we see that in the $s-$wave phase $\mathcal{S}>0$. The mode equation can be cast into $L\Omega^4 + A\Omega^2 - S=0$. This clearly has two solutions with $\Omega^2>0$ and $<0$. The $\Omega^2>0$ solution is given by the implicit in $\Omega$ equation
\beq\label{eq:asd}
\Omega^2=\frac{-A + \sqrt{A^2 + 4LS}}{2L}, ~~A>0,~L>0.
\eeq
This is the only real solution, representing the well defined BS collective mode. As we approach the $s+id$ boundary in the $s-$state, $S\rightarrow0+$ and the mass of the collective mode approaches zero (as expected). Note at this stage the following differences with the 1-band model: (1) We get a damped Leggett mode; (2) we get two BS modes where one is damped; (3) The BS mode frequency is no longer related to parameters determined at $T_c$ because of temperature dependence of the gap ratios. Depending on the magnitude of the effect (which depends on a chosen microscopic model), this should be important for quantifying the experiment with the model.

\subsubsection{Collective modes in the $s+id$ ground state}
As in the 1-band case, the $s$, $d$, amplitude and phase sectors get coupled due to $\Pi^{sd}_{12}\neq 0$. The non-zero $\Pi$'s are:
\bea\label{eq:spidpi}
\Pi^{ss}_{11,h}&=&\nu_{2D}\left[-4\int_{\theta}L_h + \left((\alpha^h_s)^2-\left(\frac{\Omega}{2\Delta_0}\right)^2\right)I^h_0(\Omega)\right],\nonumber\\
\Pi^{ss}_{11,e}&=&\nu_{2D}\left[-8L_e + 2\left((\alpha^e_s)^2-\left(\frac{\Omega}{2\Delta_0}\right)^2\right)I^e_0(\Omega)\right],\nonumber\\
\Pi^{ss}_{22,h}&=&\nu_{2D}\left[-4\int_{\theta}L_h + (\alpha^h_d)^2I^h_2(\Omega) -\left(\frac{\Omega}{2\Delta_0}\right)^2I^h_0(\Omega)\right],\nonumber\\
\Pi^{ss}_{22,e}&=&\nu_{2D}\left[-8L_e + 2(\alpha^e_d)^2I^e_2(\Omega) -2\left(\frac{\Omega}{2\Delta_0}\right)^2I^e_0(\Omega)\right],\nonumber\\
\Pi^{sd}_{12,h}&=&\nu_{2D}\left[\alpha^h_s\alpha^h_d I^h_2(\Omega)\right],\nonumber\\
\Pi^{sd}_{12,e}&=&\nu_{2D}\left[2\alpha^e_s\alpha^e_d I^e_2(\Omega)\right],\nonumber\\
\Pi^{dd}_{11,h}&=&\nu_{2D}\left[-4\int_{\theta}f^2L_h + \left((\alpha^h_s)^2-\left(\frac{\Omega}{2\Delta_0}\right)^2\right)I^h_2(\Omega)\right],\nonumber\\
\Pi^{dd}_{11,e}&=&\nu_{2D}\left[-8L_e + 2\left((\alpha^e_s)^2-\left(\frac{\Omega}{2\Delta_0}\right)^2\right)I^e_2(\Omega)\right],\nonumber\\
\Pi^{dd}_{22,h}&=&\nu_{2D}\left[-4\int_{\theta}f^2L_h + (\alpha^h_d)^2I^h_4(\Omega) -\left(\frac{\Omega}{2\Delta_0}\right)^2I^h_2(\Omega)\right],\nonumber\\
\Pi^{dd}_{22,e}&=&\nu_{2D}\left[-8L_e +2(\alpha^e_d)^2I^e_2(\Omega)-2\left(\frac{\Omega}{2\Delta_0}\right)^2I^e_2(\Omega)\right],\nonumber\\
\eea
where $L_{h,e}=\frac12  \ln\frac{2\Lambda}{\sqrt{(\Delta^{h,e}_s)^2 + f_{\bk}^2(\Delta^{h,e}_d)^2}}$.

\begin{figure}[htp]
$\begin{array}{c}
\includegraphics[width=0.8\columnwidth]{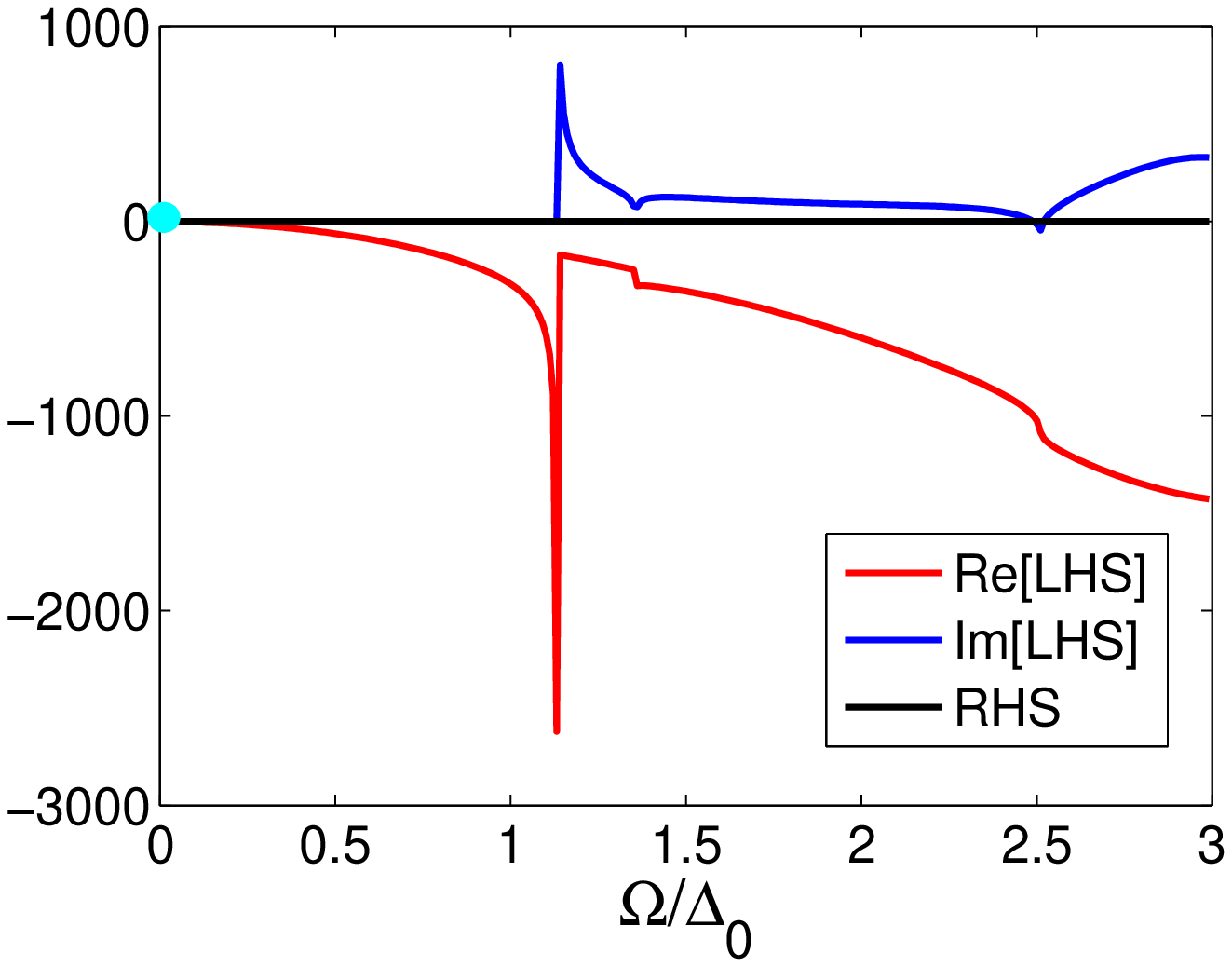}\\
\includegraphics[width=0.8\columnwidth]{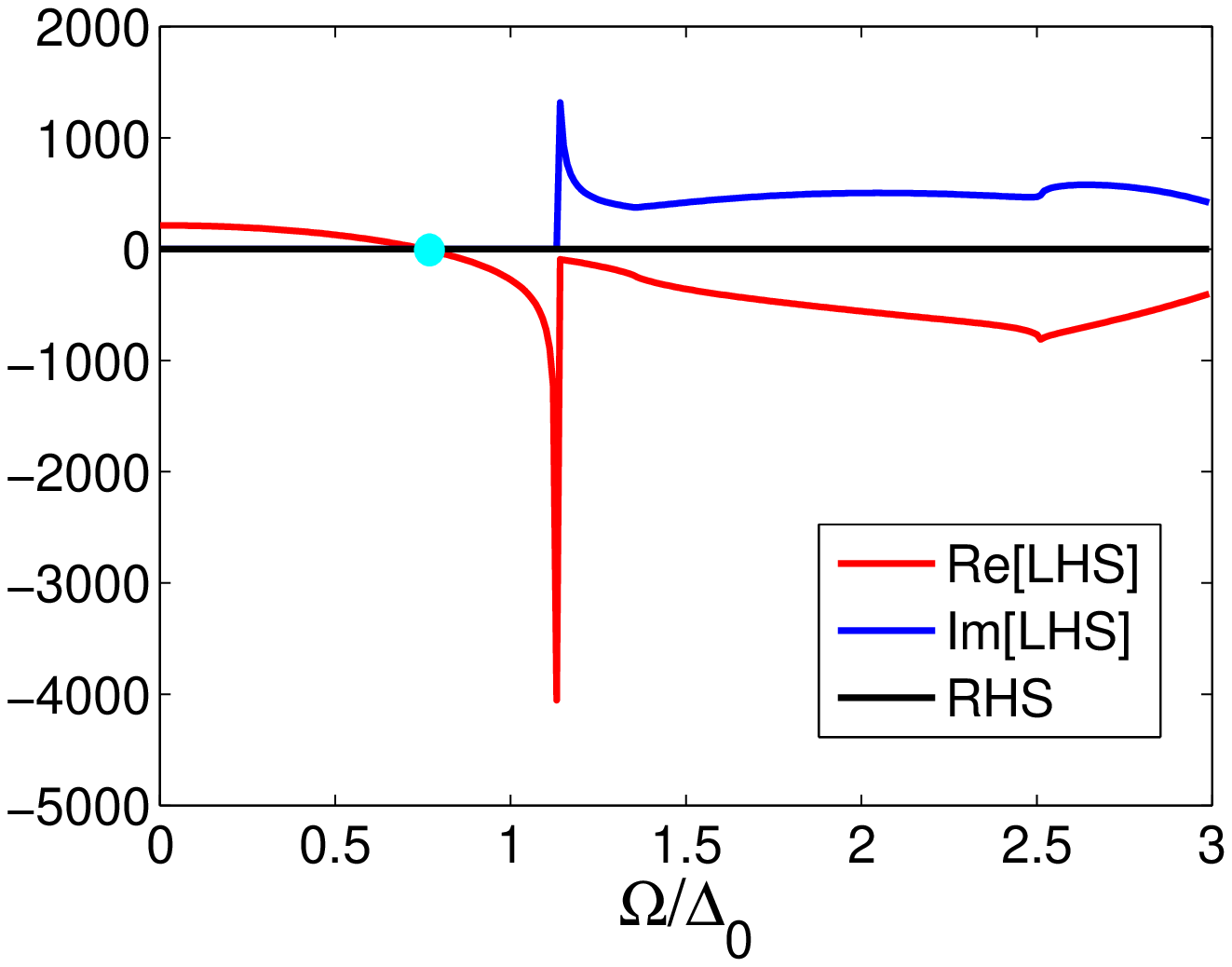}
\end{array}$\caption{
\label{fig:spidred} Solutions to Eq. (\ref{eq:det1}) (top) and its counterpart with Nambu components 1 and 2 exchanged(bottom) in the $s+id$ phase. Solution to Eq. Eq. (\ref{eq:det1}) contains the BAG mode (blue dot at $\Omega=0$) and the other equation has the mixed symmetry collective mode (blue dot at finite $\Omega$). This mode goes soft at the boundaries of the $s+id$ state. Note that in the $s+id$ state, phase and amplitude degrees of freedom are mixed in both equations. Here we used parameters $v_s=0.2$, $u_d=-0.231$, $z=1/2$.}
\end{figure}

The $T=0$ solutions suggest the following useful relations
\bea\label{eq:T0sp}
1&=&-2v_sr_s[2L_e],\nonumber\\
1&=&-\frac{v_s}{r_s}\int[2L_h],\nonumber\\
1&=&-u_d\int f^2[2L_h]-2v_dr_d[2L_e],\nonumber\\
1&=&-\frac{v_d}{r_d}\int f^2[2L_h].
\eea
We thus get two decoupled $4\times4$ sectors formed out of $22ss-21sd-11dd$ and $11ss-12sd-22dd$ parts, of the form
\beq\label{eq:det1}
\text{det}\left(
\begin{array}{cccc}
\Pi^{ss}_{22,h}&-\frac{2}{v_s}&\Pi^{sd}_{21,h}&0\\
-\frac{2}{v_s}&\Pi^{ss}_{22,e}&0&\Pi^{sd}_{21,e}\\
\Pi^{sd}_{12,h}&0&\Pi^{dd}_{11,h}&-\frac{2}{v_d}\\
0&\Pi^{sd}_{12,e}&-\frac{2}{v_d}&\Pi^{dd}_{11,e}+\frac{2u_d}{v_d^2}
\end{array}
\right)=0.
\eeq
The other equation is given by $(1~\leftrightarrow 2)$. The solutions are plotted in Fig. \ref{fig:spidred}. We obtain the BAG mode and a single mixed symmetry collective mode. The result of tracking the collective modes (with no imaginary part) as function of $u_d$ is shown in Fig. \ref{fig:2band phase}(c). There are no other damped resonances.

\section{Discussion}\label{sec:disc}
Having understood the spectrum of collective modes that can exist in a specific multiband model for a SC with competing $s$ and $d$ channels, we now discuss in some detail the practical motivation behind the choice of our model. As mentioned earlier, FeSC are complicated systems with a varying range of FS topologies and usually the minimal model depends on the particular family of FeSC of interest. We chose to study the case of $s$ to $d$ transition as it is supposed to host the exotic $s+id$ state whose detection is challenging. We propose the identification of the MSBS collective mode to detect the $s+id$ state. Heavily hole or electron doped FeSC are the best candidates for the detection of the $s+id$ state. Our model was designed keeping in mind Ba$_{1-x}$K$_x$Fe$_2$As$_2$ near $x\sim 1$. This is an interesting material which is known to be a fully gapped $s\pm$ SC at $x\sim 0.4$\cite{spm1,spm2,spm3}. It is also known to be nodal at $x\sim 1$. The symmetry of the latter compound is heavily debated: thermal conductivity\cite{dexp} data is interpreted in favor of $d$-wave which is in line with functional renormalization group prediction\cite{dtheroty}. This is strongly contradicted by the photoemission data\cite{sexp1,sexp2} which supports $s-$wave and also has theoretical support\cite{sup}. In either case, traversing from $x\sim 0.4$ to $x\sim1$ may require the system to go through exotic states like $s+id$ or $s+is'$ as the case may be. The collective modes in the $s+is'$ state \cite{SM_AVC,Benfatto}  have Leggett-like collective modes (with $s-$symmetry) that soften at the transition. We have now shown that the collective modes in the $s+id$ state have a mixed symmetry BS mode (with contributions in both $s$ and $d$ symmetry channels) which also softens at the boundary of the $s+id$ state. The onset of the $s+id$ state can thus be identified by simultaneous (and/or correlated) observation of the onset of modes in both the $s-$ and $d-$ channels of Raman spectroscopy. Experimentally, there is strong indication of a mode in the $d-$channel as claimed in some recent works.\cite{Hackl12,Hackl14} The signal in the $s-$channel is not conclusive which leaves us with two possibilities - (i) The system is still in the $s$-wave ground state, (ii) the screening effects in the $s$ channel of Raman spectroscopy washes out the mode features, in which case the intensity of the $s$ component in the MSBS must be carefully analysed. Both scenarios are unexplored theoretically as far as the material is concerned.

Its worth mentioning that the proposed $d-$wave state in the $x\sim 1$ sample has the property that SC is driven by the hole pockets\cite{dtheroty}. We thus chose to study the model where SC was driven by  $d-$wave interaction within the hole pocket.

We should warn the readers, however, that the existence of collective modes is different from actually observing them through experimental probes. One of the most promising probe to detect these modes (along with their symmetries) is electronic Raman spectroscopy. But the coupling of the modes to a Raman probe needs a special attention due to screening effects in the $s-$channel. The distribution of intensities across the $s$ and $d$ parts of the mixed symmetry collective modes is another aspect which requires careful attention. Thus, while a quantitative mapping of our results to Raman still needs more work, we are in a position to at least suggest the number and nature of the collective modes to be expected in FeSC.

Finally, although we know form the efforts of Ref. \onlinecite{WuGriffin} the possible modes in a `conventional' nodal $d-$wave state, there are other $d$-wave models like fully gapped $d-$wave state proposed for systems like alkali intercalated-FeSe\cite{alkali1,alkali2} that need exploration in terms of the behavior of collective modes and is left for a future effort. It is also worth noting that as far as mixed symmetry modes are concerned, such an analysis could be applied to systems under uniaxial strain where we generate a similar $s-d$ competition.\cite{RMF}

\section{Conclusions}\label{sec:conclu}
To summarize, we have used a simple linear response approach to study collective modes in a SC with competing $s$ and $d$ channel instabilities. We work close to the region where $s$ and $d$ channels are nearly degenerate such that the system supports an exotic $s+id$ state. We model a system that can be tuned from a pure $s$ state to pure $d$ state through the $s+id$ state. Although our goal is the description of a multiband FeSC system, we first used a 1-band model to demonstrate the simplicity of the approach. In this 1 band model, we find the massless BAG mode throughout the phase diagram in the $s-$channel. In the $d-$channel, in the $s-$wave ground state we find the BS mode and in the $s+id$ ground state find a mixed symmetry collective mode that softens at the boundaries of the $s+id$ state. These boundaries were also calculated analytically. This mixed symmetry mode, which couples amplitude and phase sectors, oscillates with both $s$ and $d$ wave components and only exists in the $\mathcal{T}$-broken $s+id$ SC state. It is interesting to note that such a mode, which we refer to
as a mixed-symmetry Bardasis-Schrieffer mode, corresponds, in a certain sense, to modes in ${\cal T}$-breaking ground states that have been discussed before,
usually in situations where the two competing interactions correspond to degenerate basis functions of a 2D representation, e.g. ``clapping modes"
in $p+ip'$ or $d+id'$ situations.\cite{YipSauls,Hirschfeld92}  More generally if there are two competing representations, a mode of this type is possible\cite{Balat}.

We generalized our approach for a multiband system and found (1) the usual BAG mode in the $s$ channel across the phase diagram; (2) damped Leggett mode between the electron and hole gaps in the $s-$wave state;  (3) 2 BS modes (where one is damped, like the Leggett mode) in the $s-$wave state; (4) A mixed symmetry collective mode in the $s+id$ state that softens at the boundaries. Based on the multiband model that was designed to minimally reproduce the qualitative effects of Ba$_{1-x}$K$_x$Fe$_2$As$_2$ near $x\sim 1$, we suggest detection of the above `symmetry-selective' collective modes. We propose that such a systematic search  can eventually settle the longstanding debate about the pairing symmetry for the $x\sim1$ samples.

{\it Acknowledgements.} We thank and T. Boehm, A. Chubukov, T. Devereaux, D. Einzel, R. M. Fernandes, R. Hackl, W. Hanke, and R. Thomale for useful discussions. PJH is grateful to P. P. W\"{o}lfle for many inspirational discussions about the nature of order parameter collective modes in superfluids and superconductors. SM is a Dirac Post-Doctoral Fellow at the National High Magnetic Field Laboratory, which is supported by the National Science Foundation via Cooperative agreement No. DMR-1157490, the State of Florida, and the U.S. Department of Energy. Both authors were supported in part by DOE DE-FG02-05ER46236.

\newpage
\begin{widetext}
\section{Appendix}
\subsection{Solving the 1-band model}
We solve the model described by Eq (\ref{eq:2}). For analytical answers will only look at $T=0$ and $T=T_c$ points to study the special points in the phase diagram of the model. We conjecture (based on continuity) that there are no other special points in the phase diagram. At $T=T_c$ ($\Delta^{s,d}\rightarrow 0$), Eq. (\ref{eq:2}) leads to
\bea\label{eq:q3}
\Delta^s&=&-u_s\Delta^s [2L_c],\nonumber\\
\Delta^d&=&-u_d\int_{\theta} f^2_{\bk}\Delta^d [2L_c].
\eea
where $\int_{\theta}$ stands for the angular integral at the Fermi surface, $L_c=\frac12\text{ln}\frac{2\gamma\Lambda}{\pi T_c}$. Using $\int_{\theta}f^2_{\bk}=1$ we see that
\beq\label{eq:4}
2L_c = \text{min}\left\{-\frac{1}{u_s},-\frac{1}{u_d}\right\};
\eeq
with the $s+id$ state being realized when $u_d=u_s$.  At $T=0$, as discussed, there are three regions. For the $s-$wave state Eq. (\ref{eq:2}) yields:
\bea\label{eq:5}
1&=&-u_s \text{ln}\frac{2\Lambda}{\Delta^s},~\Delta^d=0.
\eea
This requires $u_s<0$ (attractive) and also gives the $T=0$ value for $\Delta_s$. We shall now define
\beq\label{eq:6}
2L\equiv\text{ln}\frac{2\Lambda}{\Delta^s_0}=-\frac{1}{u_s}.
\eeq
$\Delta^s_0$ is the $s-$wave gap in the model at $T=0$ with no competing $d-$wave interaction. As the $d-$wave interaction ($u_d$) grows, $\Delta^d$ remains zero up to the point where the second equation in Eq. (\ref{eq:2}) first has a non-trivial solution, then $\Delta^d$ begins to grow and $\Delta_s$ begins to suppress. In the $s+id$ state we have
\bea\label{eq:7}
1&=&-u_s \int_{\theta}\text{ln}\frac{2\Lambda}{\sqrt{\Delta_s^2 + \Delta_d^2f^2_{\bk}}},\nonumber\\
1&=&-u_d \int_{\theta}f^2_{\bk}\text{ln}\frac{2\Lambda}{\sqrt{\Delta_s^2 + \Delta_d^2f^2_{\bk}}},
\eea
In the notation of the main text, Eq. (\ref{eq:7}) can then be cast into a simpler form-
\bea\label{eq:9}
-\frac{1}{u_s}&=& 2L_g + \int_{\theta}\text{ln}\frac{1}{\sqrt{\alpha^2_s + \eta^2\alpha_s^2f^2_{\bk}}},\nonumber\\
-\frac{1}{u_d}&=& 2L_g + \int_{\theta}f^2_{\bk}\text{ln}\frac{1}{\sqrt{\alpha^2_s + \eta^2\alpha_s^2f^2_{\bk}}},
\eea
or,
\bea\label{eq:10}
\text{ln}\alpha_s&=& -\int_{\theta}\text{ln}\sqrt{1+ \eta^2f^2_{\bk}},\nonumber\\
\frac{1}{u_s}-\frac{1}{u_d}&=& -\int_{\theta}(f^2_{\bk}-1)\text{ln}\sqrt{1 + \eta^2f^2_{\bk}}.
\eea
where $2L_g=\text{ln}\frac{2\Lambda}{\Delta^s_0}$. This system of equations gives the gap ratios $\alpha_{s,d}$ as a function of the parameter $u_d$.
\subsection{Explicit form for the 1-band mode equation}
In explicit form for the 1-band case the collective mode equation is

\beq\label{eq:finalform}
\left(
\begin{array}{cccccc}
\Pi^{ss}_{11}-2V_s^{-1}&\Pi^{ss}_{12}&\Pi^{ss}_{13}&\Pi^{sd}_{11}&\Pi^{sd}_{12}&\Pi^{sd}_{13}\\
\Pi^{ss}_{21}&\Pi^{ss}_{22}-2V_s^{-1}&\Pi^{ss}_{23}&\Pi^{sd}_{21}&\Pi^{sd}_{22}&\Pi^{sd}_{23}\\
\Pi^{ss}_{31}&\Pi^{ss}_{32}&\Pi^{ss}_{33}-V_{s,q}^{-1}&\Pi^{sd}_{31}&\Pi^{sd}_{32}&\Pi^{sd}_{33}\\
\Pi^{ds}_{11}&\Pi^{ds}_{12}&\Pi^{ds}_{13}&\Pi^{dd}_{11}-2V_d^{-1}&\Pi^{dd}_{12}&\Pi^{dd}_{13}\\
\Pi^{ds}_{21}&\Pi^{ds}_{22}&\Pi^{ds}_{23}&\Pi^{dd}_{21}&\Pi^{dd}_{22}-2V_d^{-1}&\Pi^{dd}_{23}\\
\Pi^{ds}_{31}&\Pi^{ds}_{32}&\Pi^{ds}_{33}&\Pi^{dd}_{31}&\Pi^{dd}_{32}&\Pi^{dd}_{33}-V_{d,q}^{-1}
\end{array}
\right)
\left(
\begin{array}{c}
-\delta\Delta^{R,s}\\
\delta\Delta^{I,s}\\
\delta D^s\\
-\delta\Delta^{R,d}\\
\delta\Delta^{I,d}\\
\delta D^d
\end{array}
\right)
=0
\eeq

Ignoring the density channel we get:

\beq\label{eq:finalforms}
\left(
\begin{array}{cccc}
\Pi^{ss}_{11}-2V_s^{-1}&\Pi^{ss}_{12}&\Pi^{sd}_{11}&\Pi^{sd}_{12}\\
\Pi^{ss}_{21}&\Pi^{ss}_{22}-2V_s^{-1}&\Pi^{sd}_{21}&\Pi^{sd}_{22}\\
\Pi^{ds}_{11}&\Pi^{ds}_{12}&\Pi^{dd}_{11}-2V_d^{-1}&\Pi^{dd}_{12}\\
\Pi^{ds}_{21}&\Pi^{ds}_{22}&\Pi^{dd}_{21}&\Pi^{dd}_{22}-2V_d^{-1}
\end{array}
\right)
\left(
\begin{array}{c}
-\delta\Delta^{R,s}\\
\delta\Delta^{I,s}\\
-\delta\Delta^{R,d}\\
\delta\Delta^{I,d}\\
\end{array}
\right)
=0.
\eeq

All information about collective modes are obtained by looking at the determinant of the above matrices. Depending on the relevant ground state, some of the $\Pi$'s (as discussed in the main text) are zero, thereby simplifying the matrix structure.

\subsection{Finding solutions to the 1-band collective mode equation}

Here we present the graphical solutions to some of the equations presented in the main text. Fig. \ref{fig:2} shows the $s-$wave sector solutions (amplitude and phase) to the collective mode equation in the 1-band $s-$wave ground state. Fig. \ref{fig:3} shows the $d-$wave sector solutions (amplitude and phase) to the collective mode equation in the 1-band $s-$wave ground state. Fig. \ref{fig:4} shows the collective mode solutions in the two coupled sectors in the 1-band $s+id$ ground state.

In the $s+id$ ground state, the matrix equation in Eq. (\ref{eq:1band}) becomes block diagonal in the 1s-2d and 2s-1d sectors. This mixes amplitude and phase and the $s$ and $d$ channels. The equations in the two sectors are:
\bea\label{eq:dd}
(\Pi^{ss}_{11}-\frac{2}{u_s})(\Pi^{dd}_{22}-\frac{2}{u_d})-\alpha_s^2\alpha_d^2I_2^2&=&0,\nonumber\\
(\Pi^{ss}_{22}-\frac{2}{u_s})(\Pi^{dd}_{11}-\frac{2}{u_d})-\alpha_s^2\alpha_d^2I_2^2&=&0,\nonumber\\
\eea
which simplifies to
\bea\label{eq:ds}
\left(\alpha_d^2I_4 - \left(\frac{\Omega}{2\Delta_0^s}\right)^2I_2\right)\left(\alpha_s^2-\left(\frac{\Omega}{2\Delta_0^s}\right)^2\right)I_0-\alpha_s^2\alpha_d^2I_2^2&=&0,\nonumber\\
\left(\alpha_d^2I_2 - \left(\frac{\Omega}{2\Delta_0^s}\right)^2I_0\right)\left(\alpha_s^2-\left(\frac{\Omega}{2\Delta_0^s}\right)^2\right)I_2-\alpha_s^2\alpha_d^2I_2^2&=&0.\nonumber\\
\eea

The final forms are given in Eqs. (\ref{eq:ds1}) and (\ref{eq:ds2}).

\begin{figure}[htp]
$\begin{array}{cc}
\includegraphics[width=0.4\columnwidth]{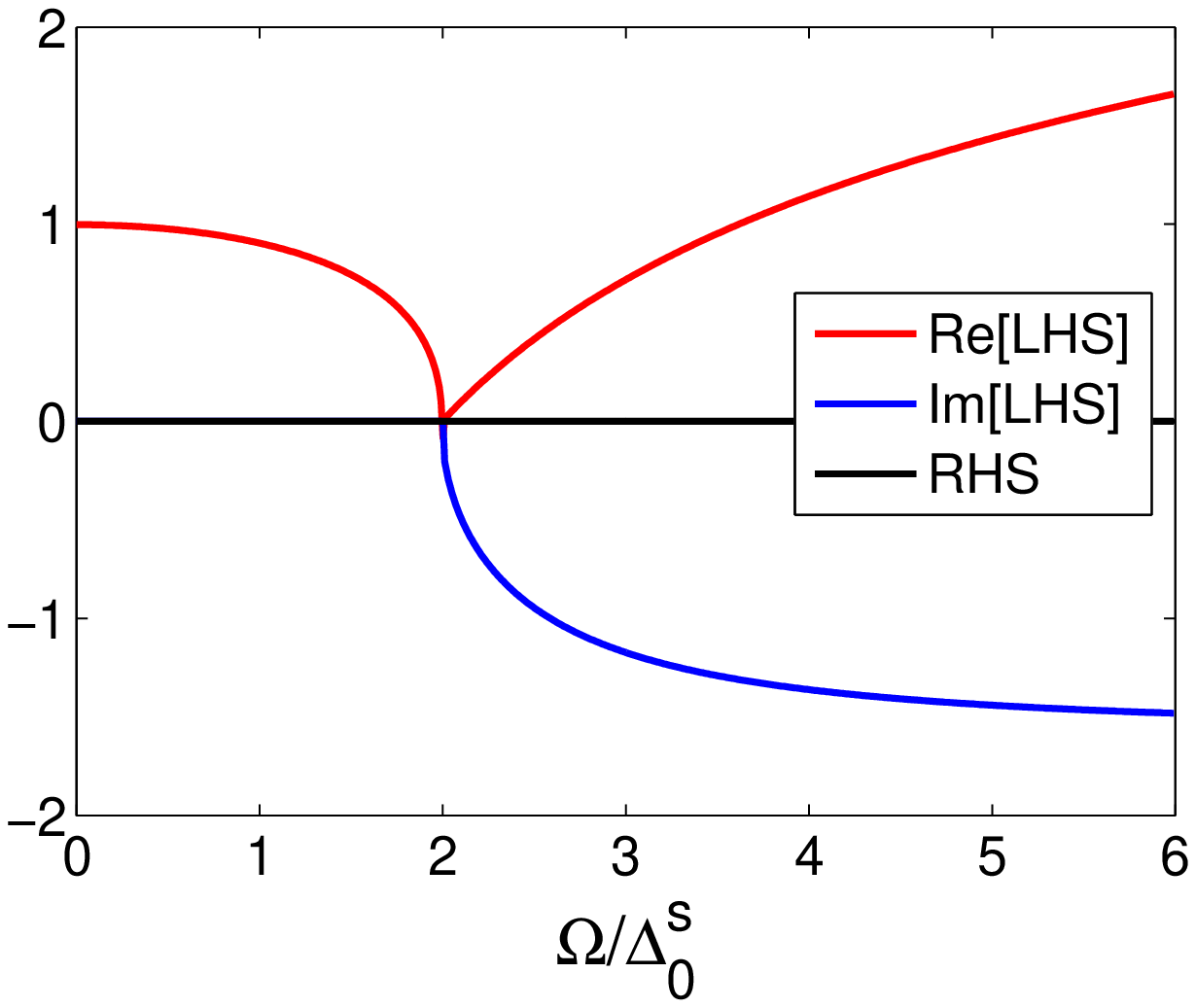}&
\includegraphics[width=0.4\columnwidth]{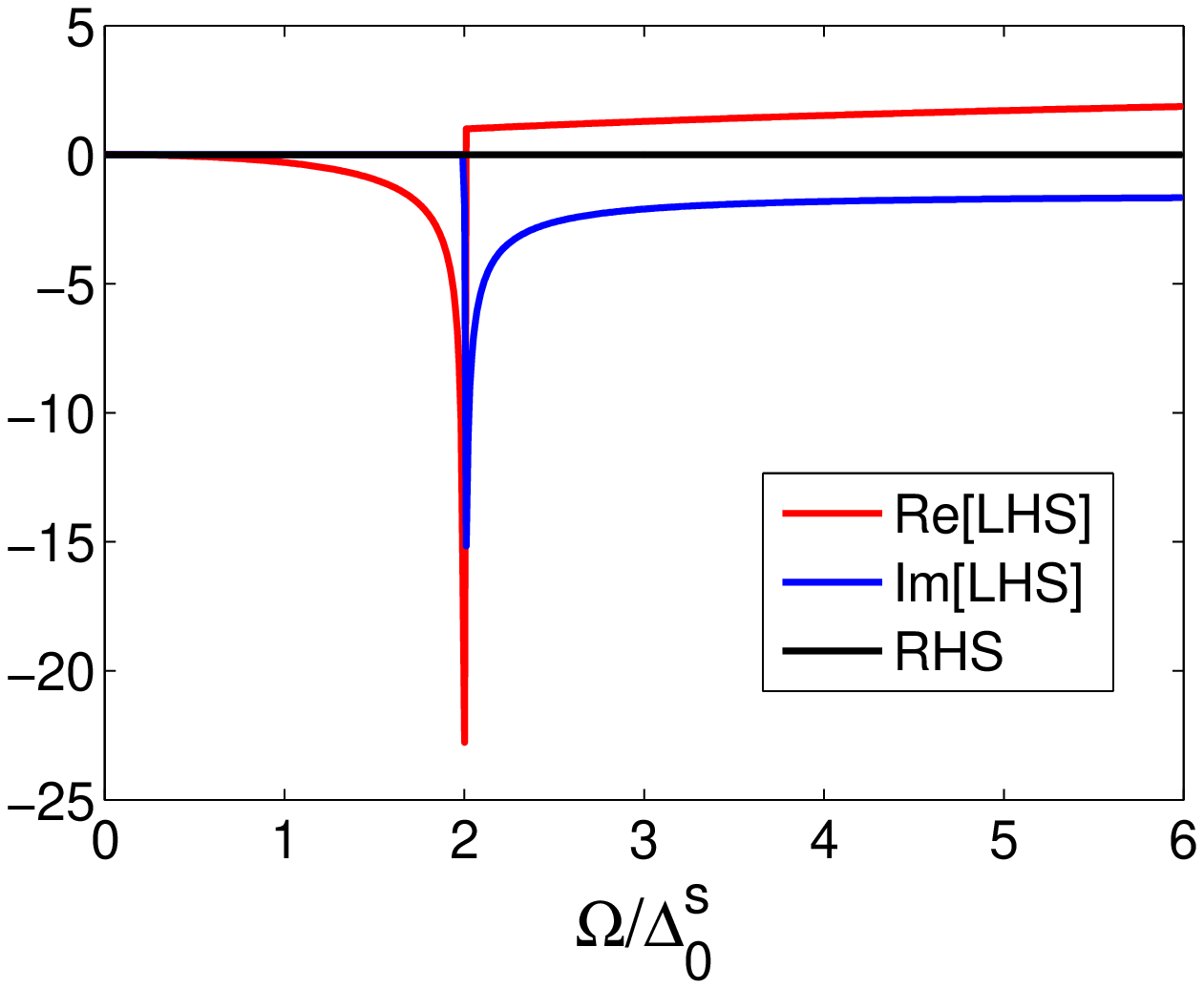}
\end{array}$\caption{
\label{fig:2} (Left) Graphical solution to the amplitude sector Eq. (\ref{eq:ssamp1}) in 1-band $s-$wave ground state with $s-$wave fluctuations. No solution for $\Omega<2\Delta_0^s$. (Right)  Graphical solution to the phase sector Eq. (\ref{eq:ssphase1}) with 1-band $s-$wave fluctuations. We get the BAG mode at $\Omega=0$.}
\end{figure}

\begin{figure}[htp]
$\begin{array}{cc}
\includegraphics[width=0.4\columnwidth]{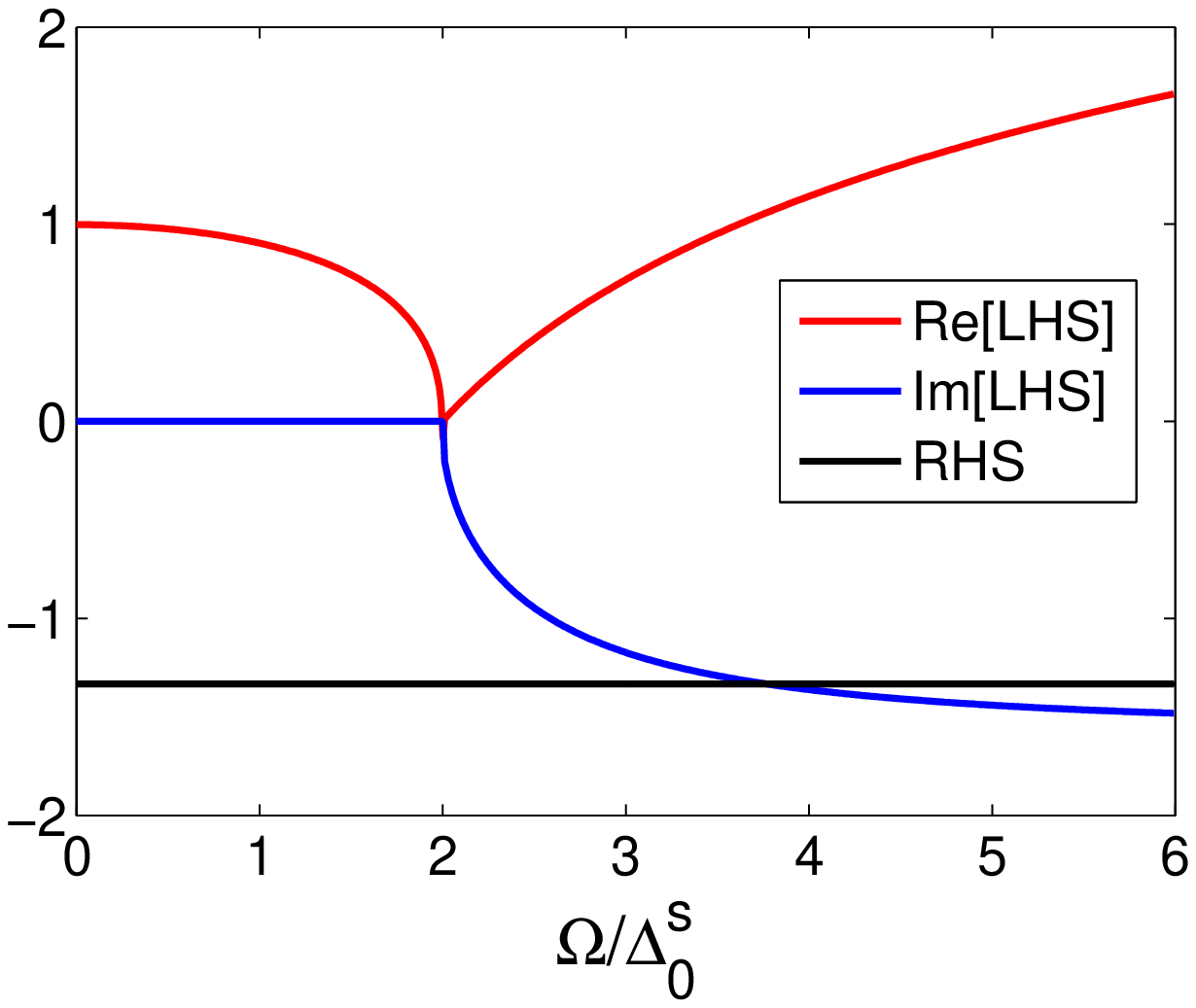}&
\includegraphics[width=0.4\columnwidth]{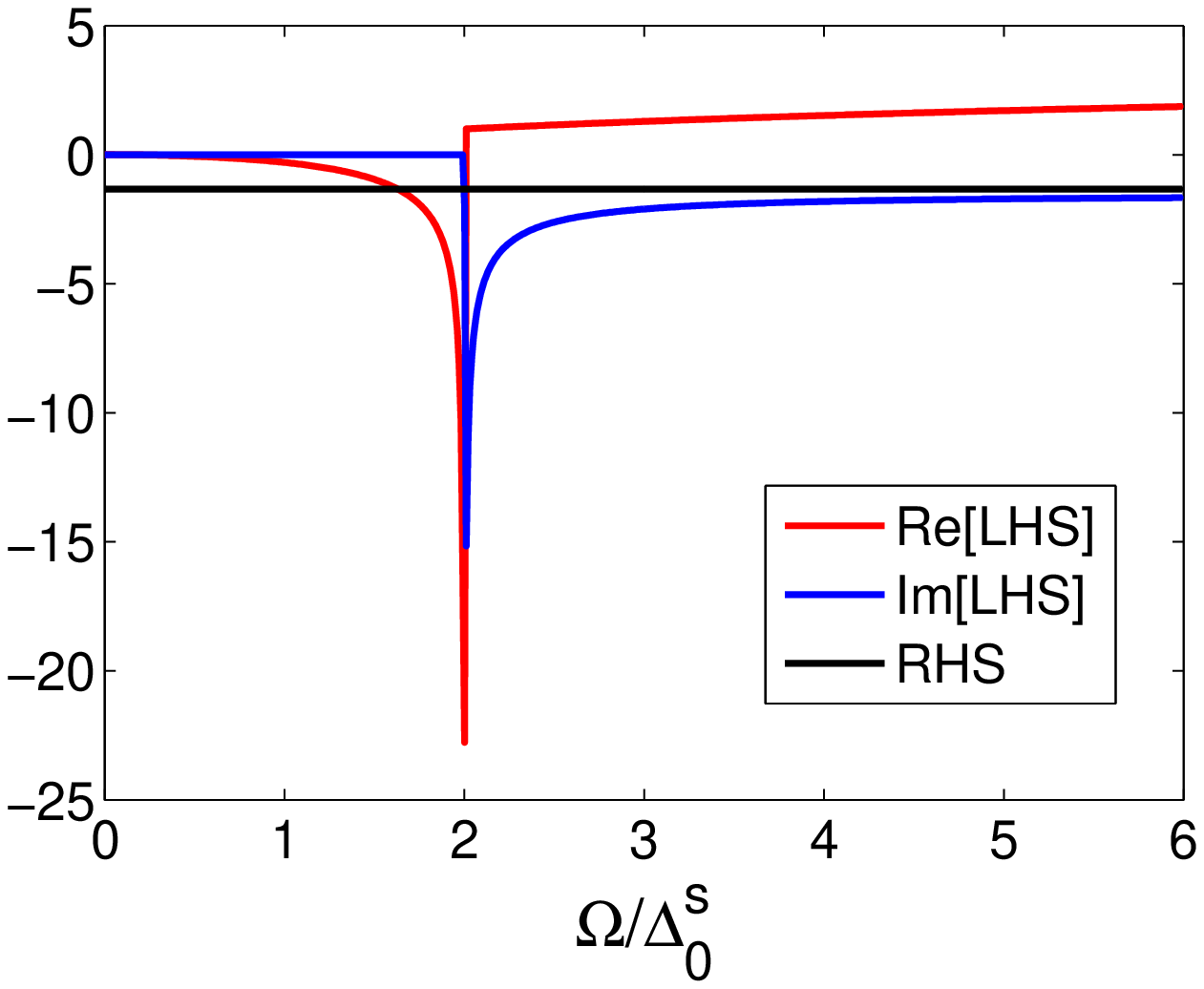}
\end{array}$\caption{
\label{fig:3} (Left) Graphical solution to the amplitude sector Eq. (\ref{eq:ssamp2}) in 1-band $s-$wave ground state with $d-$wave fluctuations. No solution for $\Omega<2\Delta_0^s$. (Right)  Graphical solution to the phase sector Eq. (\ref{eq:ssphase2}) with $d-$wave fluctuations. We get the BS mode at finite $\Omega$ that softens when $u_d = u_s$. Here, $u_s=-1$ and $u_d=-0.6$.}
\end{figure}

\begin{figure}[htp]
$\begin{array}{cc}
\includegraphics[width=0.4\columnwidth]{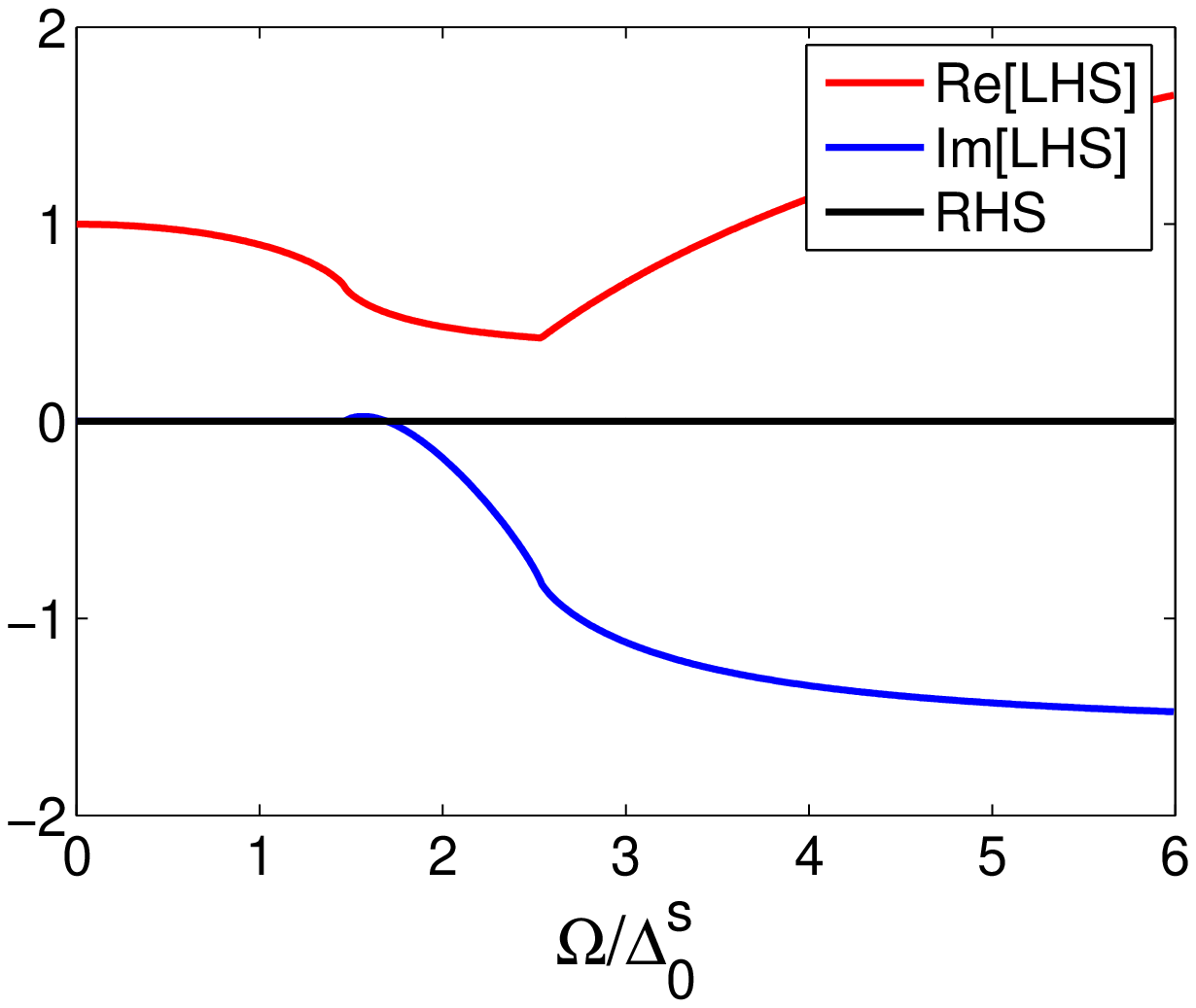}&
\includegraphics[width=0.4\columnwidth]{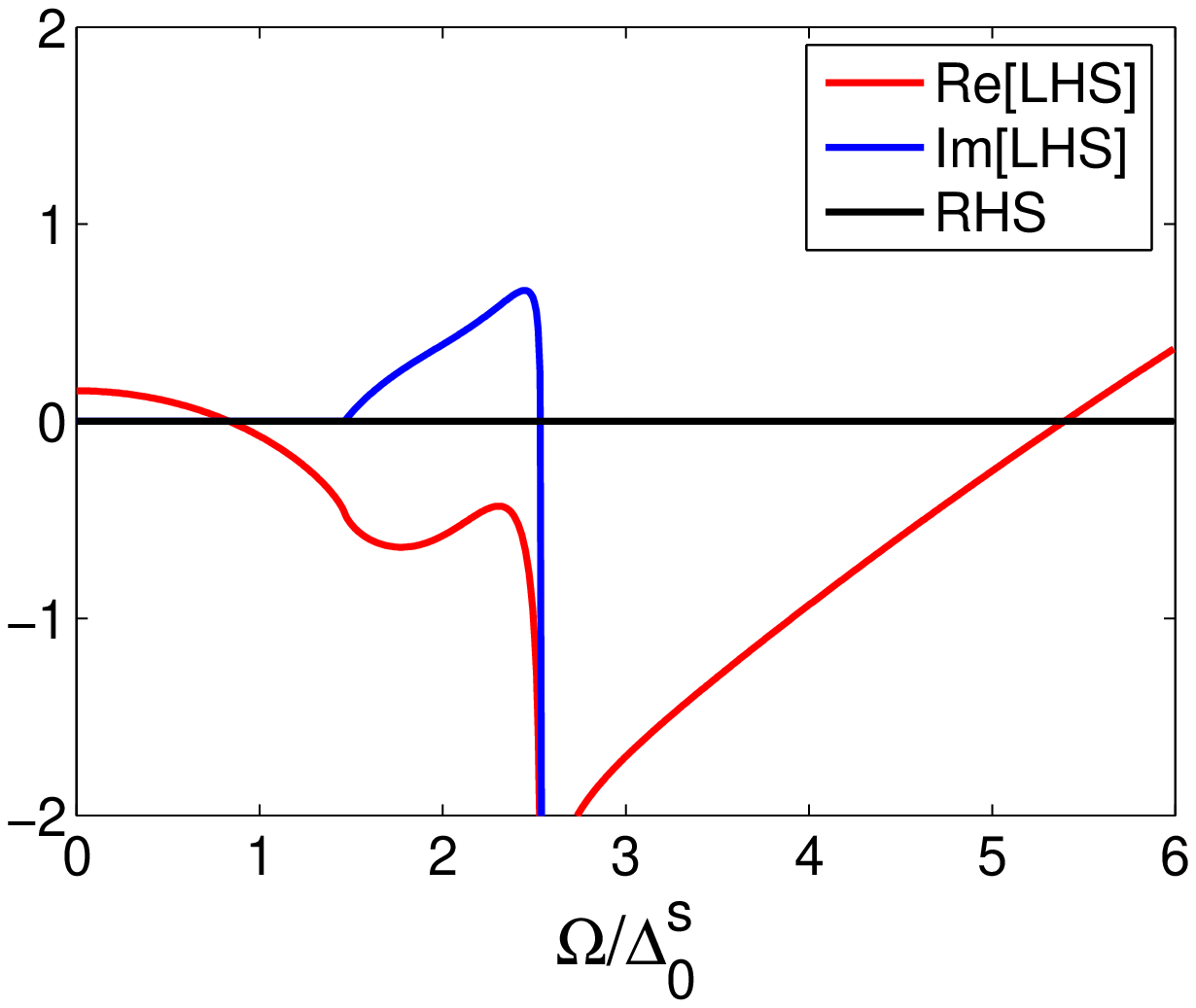}
\end{array}$\caption{
\label{fig:4} (Left) ``Amplitude" sector Eq. (\ref{eq:ds1}) -- with $\Omega^2$ removed -- in the $s+id$ ground state. No solution. (Right) ``Phase" sector Eq. (\ref{eq:ds2}) in the $s+id$ state. There is only one solution indicated by the intersection of LHS and RHS. As $u_d$ is tuned, this solution starts from 0 at the $s-$boundary and reaches $2\Delta_s$ at the other boundary-which in turn $\rightarrow 0$.}
\end{figure}

\subsubsection{The $d-$wave ground state}

\bea\label{eq:last}
\Pi_{11}^{ss}&=&\nu_{2D}\left[-\int_{\bk}\frac{1}{E}-\left(\frac{\Omega}{2}\right)^2I_0\right],\nonumber\\
\Pi_{22}^{ss}&=&\nu_{2D}\left[-\int_{\bk}\frac{1}{E}-\left(\frac{\Omega}{2}\right)^2I_0 + \alpha_d^2I_2\right],\nonumber\\
\Pi_{11}^{dd}&=&\nu_{2D}\left[-\int_{\bk}\frac{f^2}{E}-\left(\frac{\Omega}{2}\right)^2I_2\right],\nonumber\\
\Pi_{22}^{dd}&=&\nu_{2D}\left[-\int_{\bk}\frac{f^2}{E}-\left(\frac{\Omega}{2}\right)^2I_2 + \alpha_d^2I_4\right].\nonumber\\
\eea

The gap equation yields
\beq\label{eq:last2}
\frac{1}{u_d}=-\int_{\bk}\frac{f^2}{2E} = -\ln\frac{2\Lambda}{\alpha_d\Delta^s_0} +c_2.
\eeq
Also,
\beq\label{eq:last3}
\int_{\bk}\frac{1}{E}=\frac{2}{u_d}-1.
\eeq

These yield the mode equations:
\bea\label{eq:last3}
\left(\frac{\Omega}{2}\right)^2I_0&=&\frac{2}{u_d}-\frac{2}{u_s}-1,\nonumber\\
\left(\frac{\Omega}{2}\right)^2I_0-\alpha_d^2I_2&=&\frac{2}{u_d}-\frac{2}{u_s}-1,\nonumber\\
\left(\frac{\Omega}{2}\right)^2I_2&=&0,\nonumber\\
\left(\frac{\Omega}{2}\right)^2I_2-\alpha_d^2I_4&=&.\nonumber\\
\eea
It can be explicitly checked that, other than the BAG mode ($\Omega=0$), there is no solution to these equations.

\subsection{Solving the 3-pocket model}

Using the interaction form defined in Eq (\ref{eq:gap eq}) and performing $\int_{\bk}\rightarrow \nu_{2D}\int\frac{d\theta}{2\pi}\int d\e$, the gap equations at $T_c$ can be written as
\bea\label{eq: decomposed}
\Delta^h_{\bp}&=&-\left[u_df_{\bp}\int_{\theta}f_{\bk}\Delta^h_{\bk} + v_d\sqrt{\frac{m_e}{m_h}}f_{\bp}\int_{\theta}\Delta^{e_1}_{\bk} + v_s\sqrt{\frac{m_e}{m_h}}f_{\bp}\int_{\theta}\Delta^{e_1}_{\bk} + ~(e_1 \leftrightarrow e_2) \right][2L_{T_c}],\nonumber\\
\Delta^{e_1}_{\bp}&=&-\sqrt{\frac{m_h}{m_e}}\left[v_s\int_{\theta}\Delta^h + v_d\int_{\theta}f_{\bk}\Delta^h_{\bk} \right][2L_{T_c}],\nonumber\\
\Delta^{e_2}_{\bp}&=&-\sqrt{\frac{m_h}{m_e}}\left[v_s\int_{\theta}\Delta^h - v_d\int_{\theta}f_{\bk}\Delta^h_{\bk} \right][2L_{T_c}],\nonumber\\
\eea
where $\int d\e W^x_{\bk} = 2L_{T_c}\equiv \ln\frac{2\gamma \Lambda}{\pi T_c}$. In the pure $s-$wave state, we get:
\bea\label{eq:s}
\Delta^h_s&=&-v_s\sqrt{\frac{m_e}{m_h}}[2\Delta^e_s][2L_{T_c}],\nonumber\\
\Delta^{e}_s&=&-v_s\sqrt{\frac{m_h}{m_e}}\Delta^h_s[2L_{T_c}],\nonumber\\
&&\nonumber\\
\Rightarrow 2L_{T_c}&=&\frac{1}{\sqrt{2}v_s},~~~\frac{\Delta^h_s}{\Delta^{e}_s}= -\sqrt{2}\sqrt{\frac{m_e}{m_h}}
\eea
In the pure $d-$wave state, we get:
\bea\label{eq:d}
\Delta^h_s&=&-u_d\Delta^h_d[2L_{T_c}] - v_d\sqrt{\frac{m_e}{m_h}}[2\Delta^e_d][2L_{T_c}],\nonumber\\
\Delta^{e}_d&=&-v_d\sqrt{\frac{m_h}{m_e}}\Delta^h_d[2L_{T_c}],\nonumber\\
&&\nonumber\\
\Rightarrow 2L_{T_c}&=&\frac{u_d + \sqrt{u_d^2 + 8v_d^2}}{4v_d^2},~~~\frac{\Delta^h_s}{\Delta^{e}_s}= -\frac{4v_d}{u_d + \sqrt{u_d^2 + 8v_d^2}}\sqrt{\frac{m_e}{m_h}}
\eea

The $s-$wave solution will be dominant when
\beq\label{eq:deg}
\left(\frac{v_d}{v_s}\right)^2\le 1 + \frac{1}{\sqrt{2}}\left(\frac{u_d}{v_s}\right),
\eeq
with the equality reaching when the two solutions are degenerate. This will be the $s+id$ point at $T=T_c$. We shall now set $m_e=m_h$ to continue. Its going to be cumbersome otherwise.

At $T=0$, we get the following system of non-linear equations:
\bea\label{eq:T0}
\Delta^h_s&=&-v_s[2\Delta^e_s][2L_e],\nonumber\\
\Delta^e_s&=&-v_s\Delta^h_s\int_{\theta}[2L_h],\nonumber\\
\text{and}&&\nonumber\\
\Delta^h_d&=&-u_d\Delta^h_d\int_{\theta}f^2_{\bk}[2L_h]-v_d[2\Delta^e_d][2L_e],\nonumber\\
\Delta^e_d&=&-v_d\Delta^h_d\int_{\theta}f_{\bk}^2[2L_h],
\eea
where,
\bea\label{eq:L}
2L_h&=&\ln\frac{2\Lambda}{\sqrt{(\Delta^h_s)^2 + f_{\bk}^2(\Delta^h_d)^2}},\nonumber\\
2L_e&=&\ln\frac{2\Lambda}{\sqrt{(\Delta^e_s)^2 + f_{\bk}^2(\Delta^e_d)^2}}.
\eea
In the pure $s-$ or $d-$ phase we simply set the other gap to zero. At $T=0$, there are two points: where $\Delta_d\rightarrow 0$ (the $s$-wave boundary) and where $\Delta_s\rightarrow 0$ (the $d-$wave boundary).
\subsubsection{$s-$wave boundary}
At this boundary, we have
\bea\label{eq:T02}
\Delta^h_s&=&-v_s[2\Delta^e_s][2L_e],\nonumber\\
\Delta^e_s&=&-v_s\Delta^h_s[2L_h],\nonumber\\
\text{and}&&\nonumber\\
\Delta^h_d(1+u_d[2L_h])&=&-v_d[2\Delta^e_d][2L_e],\nonumber\\
\Delta^e_d&=&-v_d\Delta^h_d[2L_h].
\eea
The way to proceed is to realize that $\Delta^{h/e}_d\rightarrow 0$ but $\Delta^h_d/\Delta^e_d$ can be arbitrary. This arbitrariness can be removed by eliminating the $\Delta_d$'s from the last two equations. This will later give the constraint equation for $u_d$ to get a non-trivial $d-$wave component. The first two equations should be used to compute the $\Delta^{h/e}_s$. These yield
\bea\label{eq:T02}
\left(\frac{v_d}{v_s}\right)^2& =& 1-r_s\frac{u_d}{v_s},\nonumber\\
\text{where}~~r_s&\equiv&\frac{\Delta^e_s}{\Delta^h_d}\nonumber\\
\text{and $r_s$ satisfies}&&\nonumber\\
\frac{1}{r_s}-2r_s&=&2v_s\ln|r_s|.
\eea
Note that at $T=T_c$, $r_s=-\frac{1}{\sqrt{2}}$. At $T=0$, this will obviously depend on $v_s$.  Recall that $v_d = z u_d$ and $u_d$ is our tuning parameter for the phase diagram. This yields the critical $u_d$ to be
\beq\label{eq:cre}
u_d^{\text{crit}} = \frac{-r_s-\sqrt{r_s^2+4z^2}}{2z^2}v_s.
\eeq
And $r_s$ is to be solved as before. The result is presented in Fig. \ref{fig:2band phase}(a).

\subsubsection{$d-$wave boundary}
At this boundary, we have
\bea\label{eq:T03}
\Delta^h_s&=&-v_s[2\Delta^e_s][2L_e],\nonumber\\
\Delta^e_s&=&-v_s\Delta^h_s[2\tilde{L}_h - c_0],\nonumber\\
\text{and}&&\nonumber\\
\Delta^h_d(1+u_d[2\tilde{L}_h-c_2])&=&-v_d[2\Delta^e_d][2L_e],\nonumber\\
\Delta^e_d&=&-v_d\Delta^h_d[2L_h-c_2],
\eea
where, $c_0\equiv \int \ln |f|=-0.347$, and $c_2\equiv\int f^2\ln |f| =0.153$, such that $c_2-c_0=\frac12$. We then follow the same logic as for the $s-$wave boundary and eliminate $\Delta^{h/e}_s$ and get
\bea\label{eq:T05}
\frac{1}{2(v_s)^2}&=&[2\tilde{L}_h-c_2+\frac12][2\tilde{L}_h-\ln|r_d|],\nonumber\\
2\tilde{L}_h-c_2&=&-\frac{r_d}{zu_d},\\
\label{eq:TOF3}
\text{$r_d$ satisfies}&&\nonumber\\
\frac{1}{r_d}-2r_d&=&zu_d\left(c_2-\ln|r_d|\right),
\eea
where $z\equiv \frac{v_d}{u_d}$ and $r_d\equiv\frac{\Delta^e_d}{\Delta^h_d}$. We just find the $u_d$ for which Eq. \ref{eq:T05} is satisfied. The result is presented in Fig. \ref{fig:2band phase}(a).

\subsection{The explicit form of the 3-pocket mode equation}

\beq\label{eq:finalformB}
\left(
\begin{array}{cccccccccc}
\Pi^{ss}_{11,h}&0&\Pi^{ss}_{12,h}&0&\Pi^{ss}_{13,h}&\Pi^{sd}_{11,h}&0&\Pi^{sd}_{12,h}&0&\Pi^{sd}_{13,h}\\
0&\Pi^{ss}_{11,e}&0&\Pi^{ss}_{12,e}&\Pi^{ss}_{13,e}&0&\Pi^{sd}_{11,e}&0&\Pi^{sd}_{12,e}&\Pi^{sd}_{13,e}\\
\Pi^{ss}_{21,h}&0&\Pi^{ss}_{22,h}&0&\Pi^{ss}_{23,h}&\Pi^{sd}_{21,h}&0&\Pi^{sd}_{22,h}&0&\Pi^{sd}_{23,h}\\
0&\Pi^{ss}_{21,e}&0&\Pi^{ss}_{22,e}&\Pi^{ss}_{23,e}&0&\Pi^{sd}_{21,e}&0&\Pi^{sd}_{22,e}&\Pi^{sd}_{23,e}\\
\Pi^{ss}_{31,h}&\Pi^{ss}_{31,e}&\Pi^{ss}_{32,h}&\Pi^{ss}_{32,e}&{\blue M_{33}^{ss}}&\Pi^{sd}_{31,h}&\Pi^{sd}_{31,e}&\Pi^{sd}_{32,h}&\Pi^{sd}_{32,e}&{\blue M_{33}^{sd}}\\
\Pi^{ds}_{11,h}&0&\Pi^{ds}_{12,h}&0&\Pi^{ds}_{13,h}&\Pi^{dd}_{11,h}&0&\Pi^{dd}_{12,h}&0&\Pi^{sd}_{13,h}\\
0&\Pi^{ds}_{11,e}&0&\Pi^{ds}_{12,e}&\Pi^{ds}_{13,e}&0&\Pi^{dd}_{11,e}&0&\Pi^{dd}_{12,e}&\Pi^{sd}_{13,e}\\
\Pi^{ds}_{21,h}&0&\Pi^{ds}_{22,h}&0&\Pi^{ds}_{23,h}&\Pi^{dd}_{21,h}&0&\Pi^{dd}_{22,h}&0&\Pi^{dd}_{23,h}\\
0&\Pi^{ds}_{21,e}&0&\Pi^{ds}_{22,e}&\Pi^{ds}_{23,e}&0&\Pi^{dd}_{21,e}&0&\Pi^{dd}_{22,e}&\Pi^{dd}_{23,e}\\
\Pi^{ds}_{31,h}&\Pi^{ds}_{31,e}&\Pi^{ds}_{32,h}&\Pi^{ss}_{32,e}&{\blue M_{33}^{ds}}&\Pi^{dd}_{31,h}&\Pi^{dd}_{31,e}&\Pi^{dd}_{32,h}&\Pi^{dd}_{32,e}&{\blue M_{33}^{dd}}\\
\end{array}
\right)
\left(
\begin{array}{c}
-\delta\Delta_h^{R,s}\\
-\delta\Delta_e^{R,s}\\
\delta\Delta_h^{I,s}\\
\delta\Delta_e^{I,s}\\
\delta D^s\\
-\delta\Delta_h^{R,d}\\
-\delta\Delta_e^{R,d}\\
\delta\Delta_h^{I,d}\\
\delta\Delta_e^{I,d}\\
\delta D^d
\end{array}
\right)
\eeq
\beq\label{eq:finalformB}
-2\left(
\begin{array}{cccccccccc}
0&V_s&0&0&0&0&0&0&0&0\\
V_s&0&0&0&0&0&0&0&0&0\\
0&0&0&V_s&0&0&0&0&0&0\\
0&0&V_s&0&0&0&0&0&0&0\\
0&0&0&0&2V^s_{\bq}&0&0&0&0&0\\
0&0&0&0&0&U_d&V_d&0&0&0\\
0&0&0&0&0&V_d&0&0&0&0\\
0&0&0&0&0&0&0&U_d&V_d&0\\
0&0&0&0&0&0&0&V_d&0&0\\
0&0&0&0&0&0&0&0&0&2V^d_{\bq}\\
\end{array}
\right)^{-1}
\left(
\begin{array}{c}
-\delta\Delta_h^{R,s}\\
-\delta\Delta_e^{R,s}\\
\delta\Delta_h^{I,s}\\
\delta\Delta_e^{I,s}\\
\delta D^s\\
-\delta\Delta_h^{R,d}\\
-\delta\Delta_e^{R,d}\\
\delta\Delta_h^{I,d}\\
\delta\Delta_e^{I,d}\\
\delta D^d
\end{array}
\right)=0.
\eeq
where $M^{LL'}_{33} = \Pi^{LL'}_{33,h} + \Pi^{LL'}_{33,e}$
\end{widetext}
\end{document}